\documentclass[journal]{IEEEtran}

\usepackage{cite}
\usepackage{amsmath,amssymb,amsfonts}
\usepackage{algorithmic}
\usepackage{graphicx}
\usepackage{textcomp}
\usepackage{xcolor}
\usepackage{hyperref}
\usepackage{booktabs}
\usepackage{multirow}
\usepackage{amsmath}
\usepackage{tcolorbox}
\usepackage{adjustbox}
\usepackage{svg}
\usepackage{pifont}

\def\BibTeX{{\rm B\kern-.05em{\sc i\kern-.025em b}\kern-.08em
    T\kern-.1667em\lower.7ex\hbox{E}\kern-.125emX}}
    
\begin{document}

\title{MORTAR: Multi-turn Metamorphic Testing for LLM-based Dialogue Systems}

\author{Guoxiang (Aaron) Guo, 
        Aldeida Aleti, 
        Neelofar Neelofar,
        Chakkrit Tantithamthavorn,~\IEEEmembership{Senior Member,~IEEE,} \\
        Yuanyuan Qi, 
        and Tsong Yueh Chen,~\IEEEmembership{Fellow,~IEEE.}
   
\thanks{G. Guo, A. Aleti, C. Tantithamthavorn and Y. Qi are with the Faculty of Information Technology,
Monash University, Clayton, VIC 3800, Australia. (e-mail: {Guoxiang.Guo, Aldeida.Aleti, Chakkrit, Yuanyuan.Qi}@monash.edu)}
\thanks{Neelofar is with the School of Computing Technologies, RMIT University, Melbourne, VIC 3000, Australia (e-mail: neelofar.neelofar@rmit.edu.au).}
\thanks{T. Y. Chen is with the School of Science, Computing and Emerging Technologies, Swinburne University of Technology, Hawthorn, VIC 3122 Australia (e-mail: tychen@swin.edu.au).}
\thanks{Corresponding author: Aldeida Aleti (e-mail:  Aldeida.Aleti@monash.edu).}%
}

\maketitle
\begin{abstract}
With the widespread application of LLM-based dialogue systems in daily life, quality assurance has become more important than ever. Recent research has successfully introduced methods to identify unexpected behaviour in single-turn testing scenarios. However, multi-turn interaction is the common real-world usage of dialogue systems, yet testing methods for such interactions remain underexplored. This is largely due to the oracle problem in multi-turn testing, which continues to pose a significant challenge for dialogue system developers and researchers. In this paper, we propose MORTAR, a metamorphic multi-turn dialogue testing approach, which mitigates the test oracle problem in testing LLM-based dialogue systems. MORTAR formalises multi-turn testing for dialogue systems, and automates the generation of question-answer dialogue test cases with multiple dialogue-level perturbations and metamorphic relations (MRs). The automated MR matching mechanism provides MORTAR more flexibility and efficiency in testing. When applied to eight LLM-based dialogue systems as test objects, MORTAR reveals 51\% more bugs per test case than the best-performing single-turn metamorphic testing baseline, with 24.6\% higher precision and over 42\% of detected bugs being unique.
\end{abstract}


\section{Introduction}
Rapid development of large language models (LLMs) has led to substantial capability improvements in downstream applications. LLM-based dialogue systems e.g., the closed-source LLM-based dialogue system ChatGPT \cite{openai_chatgpt}, and the open-sourced LLM-based dialogue system Ollama\cite{ollama_docs_2025}, are now frequently used by real-world users in their daily lives. However, numerous unexpected behaviours are observed in the outputs, e.g., incorrect information and irrelevant answers \cite{chang2024surveyevalllm}. Comprehensive testing is essential to ensure the output quality of LLM-based dialogue systems. However, testing such natural language generation systems inevitably encounters the oracle problem \cite{genaist}. In software testing, the test oracle is crucial, as it determines whether the system under test behaves as expected. 

Traditionally, the development of test oracles relies on the experience of system testers and the design of test suites. Crowdsource workers also play an important role in producing dialogue datasets as test cases. Many benchmark datasets \cite{yang2018hotpotqa, hendrycks2020mmlu} can be used in reference-based testing (RBT) to simulate user input and trigger the unexpected behaviours of LLMs. The form of existing testing can be further categorised into single-turn and multi-turn. Single-turn testing involves presenting the LLM with a standalone input prompt and evaluating the quality of its immediate response. Multi-turn testing is conducted through multiple back-and-forth rounds, which simulate real-world dialogue between the user and LLM. It is worth noting that most existing testing datasets and approaches only provide single-turn testing capability \cite{chang2024surveyevalllm}. According to real-world usage, however, over 63\% of dialogues contain more than two rounds \cite{sharegpt2023}. Such a mismatch between system testing and actual usage can lead to critical issues \cite{li2024llm}.

The main challenge of multi-turn testing is the automation of test oracles. The ideal test oracle is human judgement, but it is inevitably costly and inaccessible in testing scenarios where time and resources are constrained. Numerous testing approaches rely on LLM judges to give the final verdict\cite{bai2024mt101,zheng2023llmasajudge}, while prompting an LLM to provide a pass or fail verdict inevitably introduces bias, and further harms the reliability of evaluation \cite{huang2024limitationsfinetunedjudgemodels}. Considering the gap in existing approaches, which primarily focus on single-turn interactions, and the multi-turn nature of real-world systems, this research aims to propose an automated multi-turn dialogue testing approach that decouples the test oracle from straightforward verdicts of LLM judgements. 

\begin{figure}
    \centering
    \includegraphics[width=0.9\linewidth]{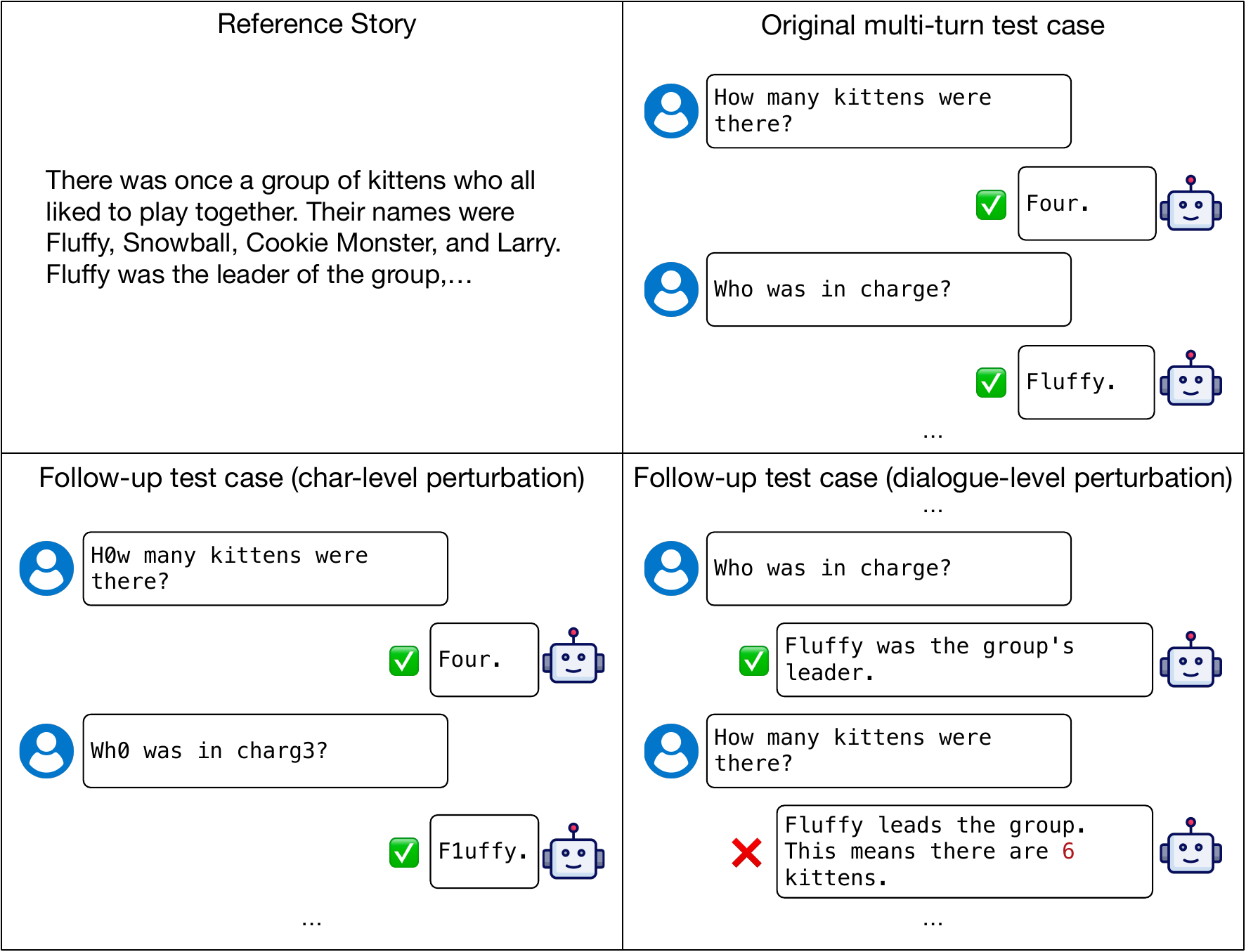}
    \caption{Testing with different test cases. The test seed is the original test case with multiple question rounds. Dialogue-level perturbation (lower right) alters the context of question and the follow-up test case trigger a failure.}
    \label{fig:intro_example}
\end{figure}

Metamorphic testing (MT) \cite{chan1998application}, a well-established software testing method, has shown the potential to alleviate the oracle problem in the testing of generative AI (GenAI) systems \cite{genaist}. Several MT-based approaches have been introduced to support single-turn testing for question-answering (QA) systems and large language models (LLMs)~\cite{chen2021qaasker, shen2022qaqa, hyun2024metal}. Perturbations in character level, word level, and sentence level show effectiveness in some testing scenarios. However, relying solely on single-turn perturbations in multi-turn dialogue testing fails to account for the essential characteristic of multi-turn interactions, as these single-turn perturbations are insufficient to verify the context dependence intrinsic to multi-turn dialogues. In Fig.~\ref{fig:intro_example}, with the original test case as a test seed, dialogue-level perturbation produces test cases with nuanced context information for the same turn-wise input and reaches significantly enlarged dialogue coverage.

In this paper, we focus on general multi-turn QA dialogues, in which each question has a single factual answer, and the multi-turn interactions are used to clarify or decompose the information needed to reach that answer \cite{reddy2019coqa}. Under this scope, we propose MORTAR, a metamorphic multi-turn dialogue testing approach, which mitigates the oracle problem in the testing of LLM-based dialogue systems. Following the definition of existing works \cite{wang2024kgit, shen2022qaqa}, we regard the unexpected output of system under test as a bug, i.e., violation of MR in MT. Using the open-domain multi-turn dialogue dataset \cite{reddy2019coqa}, MORTAR generates follow-up test cases to implement multi-turn MT for LLM-based dialogue systems. MORTAR formalises four MRs and proposes seven dialogue-level perturbations, using the multi-turn MT framework to generate follow-up test cases and reveal bugs in dialogue systems. The matching of MR is not conducted manually, but can be automated with an equivalent context check process, such that MORTAR is fully automated. According to the experimental results on multiple LLM-based dialogue systems, MORTAR achieves higher effectiveness in testing. Particularly, MORTAR reveals 164\% more bugs than the most effective single-turn MT baseline \cite{hyun2024metal}. Regarding the quality of bugs, MORTAR achieves better performance in terms of diversity, precision and uniqueness. The ablation studies verify the contribution of MRs and perturbations to the overall effectiveness. Implementation of MORTAR is made public\textsuperscript{\ref{fn:repo}}. 

\footnotetext[1]{\label{fn:repo}\url{https://github.com/Guoxiang365/MORTAR} and \url{https://zenodo.org/records/19556799}}

The main contributions of this paper are as follows:
\begin{enumerate}
\item We formalise the framework of multi-turn metamorphic testing, which serves as the foundation of MORTAR.
\item Three basic and four derivative dialogue-level perturbations, and four MRs are proposed to reveal bugs in LLM-based dialogue systems.
\item To enable full automation of the testing process, the effect of each perturbation on the dialogue context is identified by an equivalent context check, which drives the selection of the proper MR for violation detection.
\item MORTAR successfully reveals more bugs in different dialogue systems regarding quantity, diversity, precision and uniqueness. Experiment results show that MORTAR outperforms the most effective single-turn metamorphic testing approach in identifying unique bugs.
\end{enumerate}

The rest of this paper is structured as follows: Section~\ref{sec:BG} introduces the background and motivation. Section \ref{sec:approach} elaborates on the details of MORTAR. Section~\ref{sec:exp} describes the settings of experiments. Section~\ref{sec:results} presents the experiment results analysis and discussion. Section~\ref{sec:validity} reports threats to validity. Finally, Section~\ref{sec:LR} introduces relate works and Section~\ref{sec:conclusion} concludes and discusses potential future works.

\section{Background and motivation}
\label{sec:BG}
This section provides the necessary background to contextualise our work, covering key areas including the current landscape of testing LLM-based dialogue systems, the role of MT in addressing oracle challenges, and the critical distinction between single-turn and multi-turn testing approaches.

\subsection{Testing and evaluation of LLM-based dialogue systems}
Dialogue systems are now broadly applied in many domains, especially after the recent rapid development of LLMs \cite{fan2020DSsurvey, guo2023LLMEvalSurvey}. To ensure the quality of natural language generation applications, multiple datasets and evaluation methods emerged to test the dialogue system from many different aspects \cite{hendrycks2020mmlu, mehri2020dialoglue, li2024halluvault}. The test datasets typically comprise test input utterances and the expected outputs. Given a certain dialogue context and an input utterance, a bug detection is performed to judge if the output of the dialogue system meets the correctness criteria. If not, the detection will be recorded as positive to indicate a bug in the dialogue system. It is the common practice to test using such a reference-based scheme \cite{li-etal-2024-leveraging-large}.

The testing of dialogue systems can be broadly categorised into single-turn and multi-turn. In single-turn testing, the system is provided with one input utterance at a time, and its response is evaluated against expected behaviour \cite{hyun2024metal}. In contrast, multi-turn testing involves a sequence of inputs and outputs, simulating real conversational flows where the system must maintain context, coherence, and consistency across multiple exchanges \cite{ou2023dialogbench}. Most existing approaches for testing LLM-based dialogue systems are single-turn. 

Single-turn testing faces several well-known challenges. One major issue is the difficulty and high cost associated with acquiring reliable testing datasets~\cite{chen2021qaasker}. Additionally, there are growing concerns about potential data contamination in existing LLM training datasets~\cite{mirzadeh2024gsm}, raising questions about the continued effectiveness of tests based on such datasets. These challenges are amplified in multi-turn testing. A multi-turn test case is a series of simulated user inputs of multiple utterances with corresponding expectations on output. High-quality multi-turn dialogue test datasets are significantly less available than their single-turn counterparts~\cite{sun2024parrot}. A recent survey~\cite{chang2024surveyevalllm} found that only 2 out of 46 evaluations focus on multi-turn interactions. 

To mitigate the oracle problem, prompting LLMs to generate test datasets or score the outputs of system-under-test is adopted in an increasing number of studies. Recent multi-turn testing focuses on using LLMs to generate test cases and invite LLMs to function as judges to score the output of LLM-based dialogue systems \cite{zheng2023llmasajudge, kwan2024mteval, wang2023mint, bai2024mt101, duan2024botchat}. The concern of using LLMs to generate test cases for LLM-based dialogue system testing is that LLMs are trained on massive datasets to converge and act normally, but testing requires test cases to be unfamiliar and diversified. Using LLM to generate test cases and evaluate the outputs faces issues such as the lack of diversity of test cases, hallucination, and bias \cite{huang2024limitationsfinetunedjudgemodels}. Besides, these approaches generally focus on specific aspects, e.g., human-likeness\cite{duan2024botchat} or dialogue capability benchmark \cite{kwan2024mteval, bai2024mt101}, and omit the pass/fail test oracle and criteria on the same test seeds, resulting in limited comparability with broader dialogue system testing methods. These factors lead to the unsatisfactory multi-turn testing of LLM-based dialogue systems. There exists a notable gap between current testing approaches and expectations of LLM-based dialogue systems. This paper introduces a multi-turn dialogue testing approach based on metamorphic testing. The usage of LLM in MORTAR is limited to constrained natural language processing task rather than adopting straightforward LLM judges to give end-to-end verdicts.

\subsection{Metamorphic testing}
RBT relies on dataset annotation as the test oracle to judge the quality of output, and it is not feasible if the test oracle is inaccessible or no longer effective. MT was first proposed by Chen et al. in 1998 \cite{chan1998application} to test numerical programs without reliance on test oracles. Over the past development, MT has been widely used to mitigate the oracle problem in software testing, especially for those low-resource testing scenarios with limited availability of testing datasets \cite{segura2016surveyMT}. The fundamental idea of MT is to formalise MRs and generate follow-up test cases from the \textit{test seeds}, then check if the output of the system with follow-up input violates the MRs \cite{chen2020metamorphic}. The violation will be regarded as a system bug.

According to recent research, MT has been widely used in both testing traditional software systems and machine learning software systems \cite{chen2018metamorphic, zhang2020machine}. Test seed is the original test case that can be used to produce follow-up test cases. In testing dialogue systems, MT is promising in exploring more unique bugs with limited testing seeds~\cite{chen2021qaasker, shen2022qaqa, hyun2024metal}. However, there is no best practice for applying MRs in multi-turn dialogue testing. Previous testing is generally carried out in a single-turn manner where the perturbations applied to the dialogue system are low-level, including character-level, word-level, and sentence-level~\cite{liu2021dialtest, tu2021metamorphic}.

Considerable single-turn MT methods are not applicable to multi-turn scenarios. For example, QAQA \cite{shen2022qaqa} is one of the latest single-turn MT approaches for dialogue systems. It splits the testing dataset into two parts and uses one as a searching space to find applicable perturbation elements and adds them to the target question. In multi-turn testing, QAQA is not feasible as: a) many questions in dialogue are too short to accurately calculate semantic of an individual short sentence that is unclear, which makes searching for related content infeasible; b) QAQA expects the candidate utterance from searching space to be self-contained and feasible to be injected into the target question, yet considerable questions in multi-turn dialogue are context-sensitive. Thus they cannot be inserted into other dialogues, and the dataset itself can not be used as a reliable searching space.

Another major difference between single-turn and multi-turn MT is that, for single-turn MT, perturbations are generally predefined to be either semantic-preserving or semantic-altering, and they will be manually matched with corresponding MR and then be used in MT \cite{chen2021qaasker, shen2022qaqa, liu2021dialtest, hyun2024metal}. In multi-turn MT, the effect of dialogue-level perturbations might need additional processing to judge the actual effect on the test seeds. It is necessary yet difficult to tell whether a target question in the perturbed test case is given sufficient information from context to be answerable. This brings an additional challenge to multi-turn MT. Dialogue-level perturbation and MR have not been proposed in the prior work for its intrinsic difficulty in modelling and implementation. 

In summary, realising multi-turn MT for dialogue system remains challenging for the following reasons:
\begin{itemize}
    \item Relatively limited availability of multi-turn test seeds
    \item Absence of formalised multi-turn MRs
    \item Lack of dialogue-level perturbations
    \item Difficulty in test automation
\end{itemize}

To our knowledge, we are the first to systematically introduce dialogue-level MRs with implementation of effective and automated metamorphic test case generation pipelines and MR violation detection to mitigate the oracle problem in multi-turn dialogue system testing.

\section{Approach}
\label{sec:approach}
In this section, we introduce the framework of MORTAR, the formalised MRs and perturbations, and a feasible design of test automation. 

\subsection{Overview of MORTAR}

\begin{figure}[]
  \centering
  \includegraphics[width=1.0\linewidth]{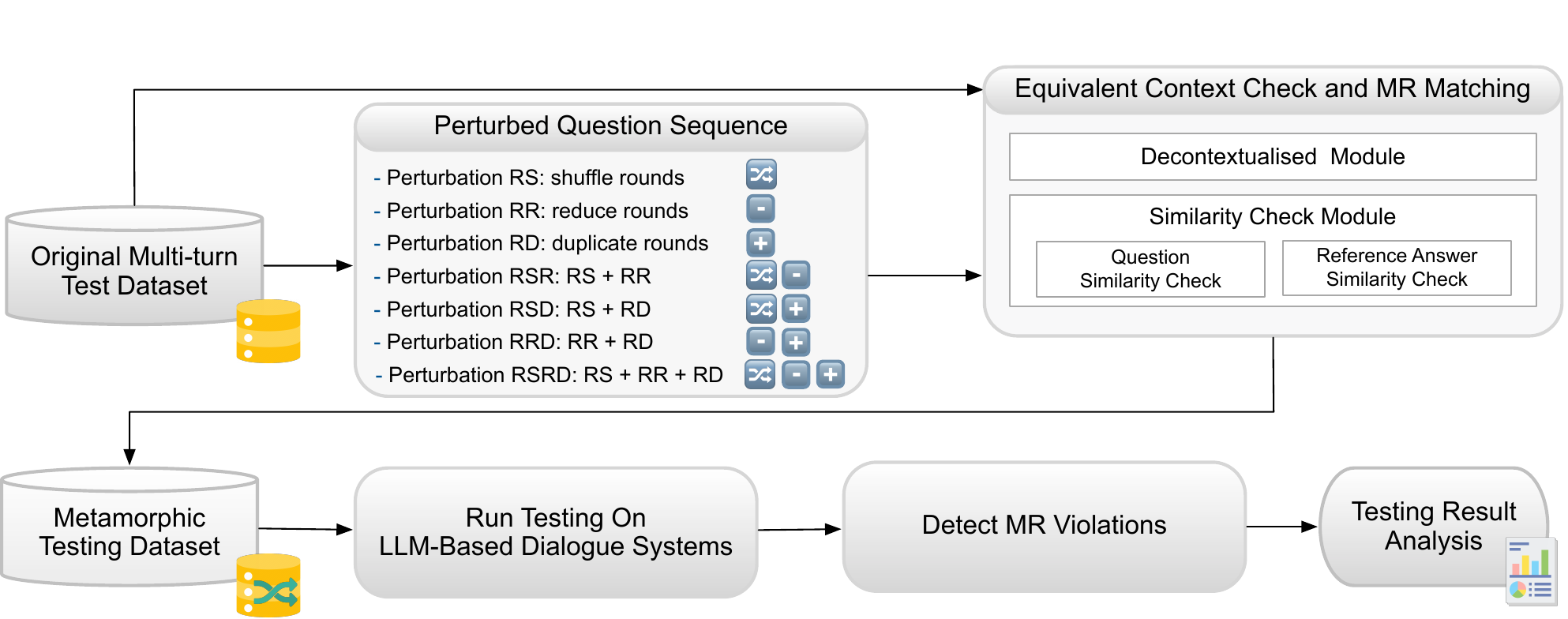}
  \caption{An overview workflow of MORTAR.}
  \label{fig:OverallFolwChart}
\end{figure}

As shown in Fig. \ref{fig:OverallFolwChart}, given the original dialogue dataset, MORTAR first generates the perturbed follow-up test input with three fundamental and four derivative dialogue-level perturbations (Section~\ref{sec:dialogue-level-perturbations}). Second, the equivalent context check verifies if the context of each question in the perturbed test input is equivalent to the original context. In each input turn, the equivalent context check result will be used to match each question with the proper MR for bug detection (Section~\ref{sec:mr-matching}). Finally, after running follow-up tests on LLM-based dialogue systems, MORTAR detects violations of MRs in dialogue system outputs. Ideally, if the dialogue systems can pass the RBT with original test cases, they are expected to pass MORTAR as well. In this way, the original test cases are reused as the test seeds, and MORTAR generate metamorphic test cases to further test the dialogue systems.

\subsection{Formalising Multi-turn Metamorphic Testing Framework}
\label{sec:MTMT}
In MORTAR, we focus on the open-domain general multi-turn QA testing scenarios where the correct answer to each question is unique under certain context. Each piece of dialogue $D$, it is composed of $n$ question answer pairs:
\begin{align}
    D= [(q_{1}, a_{1}),(q_{2}, a_{2}),\dots,(q_{n}, a_{n})]
\end{align}
where the ordering of $(q_i,a_i)$ matters, and all questions $q_i$ and answers $a_i$ in $D$ can be regrouped as two lists, $Q$ and $A$:
\begin{align}
    Q&= [q_{1},q_{2},\dots,q_{n}], \\
    A&= [a_{1},a_{2},\dots,a_{n}].
\end{align}

We use subscripts on lists to denote the initial subsequence:
\begin{align}
    D_i & : = [(q_{1}, a_{1}),\dots,(q_i, a_i)],\\
    Q_i &:= [q_1,\dots,q_i],\\
    A_i &:= [a_1,\dots,a_i],
\end{align}
where $i \in \{1,\dots, n\}$, $D_i$ is the first $i$ elements of $D$, similarly for $Q_i$ and $A_i$, such that $D_{i-1}$ can be regarded as context information when the dialogue system receives $q_i$, and the context of $q_1$ is empty, i.e. $D_0=[]$. An ideal dialogue system, $IDS(\cdot)$, exhibits the following behaviour:
\begin{align} 
IDS(D_{i-1}, q_i) &\rightarrow a_i \label{eq:oracle_contex}.
\end{align}

Equation \eqref{eq:oracle_contex} represents questions that require contextual information, which is common in multi-turn dialogue. The user usually refers to information or requirements given in context with pronouns or other types of ellipsis \cite{zhang2020ellipsis} for simplicity and convenience. However, if the contextual dependence is disrupted by perturbations and the critical information is missing in context, the target question shall not be answered with the original answer. In other cases, the target question remains context equivalent if: a) the redundant or unrelated rounds to the target question are removed by perturbations; or b) the target question can be answered independently without reliance on the context, e.g., the first question in original dialogue ($q_1$), which is usually information-complete and can be answered under empty context.

To take a close look at the context dependency in Equation \eqref{eq:oracle_contex}, the context of $q_i$ in input, $D_{i-1}$, is composed of the previous context $D_{i-2}$, the question $q_{i-1}$, and the answer $a_{i-1}$. The answer $a_{i-1}$ in the previous round is further produced by earlier round:
\begin{align}
    IDS(D_{i-2}, q_{i-1}) \rightarrow a_{i-1} \label{eq:oracle_contex_previous}.
\end{align}

Such that the information in $a_{i-1}$ has been implied by question $q_{i-1}$ and its context $D_{i-2}$. Equation \eqref{eq:oracle_contex} can be rewritten as:
\begin{align}
    IDS([D_{i-2}, q_{i-1}], q_i) &\rightarrow a_i \label{eq:oracle_contex_update_1}.
\end{align}

Similarly, since all previous answers are dependent on the sequence of questions in the corresponding context, Equation \eqref{eq:oracle_contex_update_1} can be expressed as:
\begin{align}
    IDS([q_{1}, \dots, q_{i-1}],q_i) \rightarrow a_i.
\end{align}

Or, after the definition of $Q_{i-1}$, as follows: 
\begin{align}
    IDS(Q_{i-1}, q_i) &\rightarrow a_i \label{eq:oracle_q_context}.
\end{align}

It is intuitive that the answer can be regarded as implied in the question sequence of context.

Conducting RBT on a dialogue system $DS(\cdot)$ with the original test case, is operated as feeding the dialogue system with question sequence $Q$ then verifying the quality of each output. Given the context question sequence $Q_{i-1}$ and the target question $q_i$, the dialogue system generates responses $o_i$:
\begin{align}
    DS(Q_{i-1}, q_i) \rightarrow o_i.
\end{align}

It is expected that the dialogue system under test generate similar answers to the IDS when fed with the same inputs:
\begin{align}
\forall{i \in \{1,\dots,n\}} \big(\Delta(o_i, a_i) < \epsilon\big)
\end{align}
that is:
\begin{align}
\forall{i \!\in\! \{1,\!\dots\!,n\}}\big(\Delta( DS(Q_{i-1}, q_i), IDS(Q_{i-1}, q_i)) \!<\! \epsilon\big)
\end{align}
where $\Delta(\cdot)$ measures the difference between two natural language sentences. In RBT, the difference is expected to be smaller than a threshold $\epsilon$. Otherwise, a bug will be reported under the current test case. One dialogue test case might be capable of revealing multiple bugs in different rounds, which will be recognised as different bugs revealed by this particular test case. 

The follow-up test case is defined as $Q'$:
\begin{equation}
    Q' = [q'_1,q'_2,\dots,q'_{n'}]
\end{equation}
where $n'\geq 2$, and $n'$ may not be equal to $n$. Practically, $Q'$ can be constructed from $Q$ through perturbation operations, e.g., element permutation, element deletion, element addition, and compositions of these operations, that is:
\small
\begin{equation}
\!\forall{i} \! \in \! \{1,\dots,{n'}\}  \exists j \! \in \! \{1,\dots,n\}\big( q'_i\!=\!q_j\text{ and } q_j \text{ appears in } Q\big)
\end{equation}
\normalsize

After perturbation, each $q'_i$ in $Q'$ may have different context when compared to their identical question $q_j$ in $Q$. If $q'_i$ and $q_j$ deliver the same answer, then we regard $Q'_{i-1}$ and $Q_{j-1}$ as equivalent context and denote them as $Q'_{i-1} \equiv Q_{j-1}.$ With the equivalent context, it is expected that the output from dialogue system, $o'_i$, should be similarly close to the answer of IDS $a_{j}$ when compared with $o_{j}$. That is, we have:
\small
\begin{equation}
\forall i \!\in\! \{1,\!\dots\!,n'\}\big(\Delta(o'_i,a_j) \!\approx\! \Delta(o_j,a_j)\big)\! \text{ iff } \!\big(Q'_{i-1}\!\equiv\!Q_{j-1}\big)
\label{eq:root_mr}
\end{equation}
\normalsize

In practice, if $a_j$ is inaccessible and verification of $ Q'_{i-1}\equiv Q_{j-1}$ is feasible, we have the following:
\begin{align}
\forall i\!\in\!\{1,\dots,n'\}\big(\Delta(o'_i,o_j) \leq \epsilon\big) \text{ iff } \big(Q'_{i-1}\equiv Q_{j-1}\big)
\label{eq:root_mr_prac}
\end{align}

where $\epsilon$ is the similarity thresholds. If the MR is violated in follow-up testing, a bug will be revealed using target question $q_j$ and test seed $D$.

\subsection{Dialogue-level Perturbations in MORTAR}
\label{sec:dialogue-level-perturbations}

Unlike previous perturbations that operate at the single-turn \cite{hyun2024metal,liu2021dialtest,shen2022qaqa}, MORTAR employs dialogue-level perturbations. Existing perturbations in single-turn MT can only affect the target question utterance. In contrast, dialogue-level operations are more substantial, as they can alter the conversational context, thereby influencing the answerability of the target question in perturbed test cases. This shift enables the evaluation of language models in more complex, multi-turn settings. Although these modifications are broader in scope, they are still considered perturbations - systematic variations introduced to assess the quality of dialogue systems.

Given the original question sequence, $Q=[q_{1}, q_{2}, \dots, q_{n}]$, the dialogue-level perturbation $r$ is implemented with $P^r(\cdot)$ that produces a new sequence with $n^r$ questions: 
\begin{align}
    &Q^r = P^r(Q)\\
    &Q^r = [q^r_{1}, q^r_{2}, \dots, q^r_{n^r}].
\end{align}

Let $\delta^r$ be the change of number of rounds after applying $r$: 
\begin{align}
\delta^r = \lvert n^r - n\rvert.
\end{align}

Each question $q^r_i, i\in\{1,\dots,n^r\}$ in $Q^r$ can be found in $Q$, but the round that $q^r_i$ appears in $Q^r$ might be different from the round in which the identical question $q_j$ appears in $Q$, i.e. $i$ might not equal to $j$, thus $q^r_i$ might have an inequivalent context when compared to $q_j$. Such situations will need additional judgment to tell if the condition in Equation \eqref{eq:root_mr} is met. This is one of the key features of dialogue-level perturbations, whose effect is variable for each utterance in the test seed. However, this characteristic enables them to generate rarely-seen test cases for dialogue systems when compared with single-turn perturbations, and requires the system under test to adapt to the actual dialogue and output wisely, indicating higher potential effectiveness of testing.

In MORTAR, three fundamental and four derivative dialogue-level perturbations are proposed to produce new question sequences.

\begin{itemize}
    \item \textbf{Perturbation RS}: round shuffle:
    \small
    \begin{equation}
        Q^\text{RS} = P^\text{RS}(Q) = [q_{i'} | i' = \varphi_\text{RS}(i), i=1,\dots, n]
        \label{eq:p1rs}
    \end{equation}
    \normalsize
    where $\varphi_\text{RS}(i)$ is a bijective mapping from the original round index to the shuffled round index. All original questions are reordered and form the perturbed question sequence. Perturbation RS can be regarded as element permutation of $Q$.

    \item \textbf{Perturbation RR}: round reduction:
    \small
    \begin{equation}
    Q^{\text{RR}} = P^\text{RR}(Q) = [ q_{i} | \mathbb{I}(i) = 1, i = 1, \dots, n]
    \label{eq:p2rr}
    \end{equation}
    \normalsize
    where $\mathbb{I} : \{1, \dots, n\} \rightarrow \{0,1\}$ is a randomly sampled mask function, $\sum_{i=1}^{n} \mathbb{I}(i) = n - \delta^{\text{RR}}$, and each element in $Q^{\text{RR}}$ is from $Q$. That is, $\delta^{\text{RR}}$ questions are randomly removed from the original sequence, and the rest form the perturbed question sequence. Perturbation RR can be regarded as element deletion of $Q$.
    
    \item \textbf{Perturbation RD}: round duplication:
    \begin{equation}
    Q^\text{RD} = P^\text{RD}(Q)= \text{Insert}(Q, X),
    \label{eq:p3rd}
    \end{equation}
    where $X$ is a random subset of $Q$ with $\delta^{\text{RD}}$ elements, the function $\text{Insert}(\cdot)$ iteratively inserts element from $X$ into a random position between two elements or the head or tail position, finally forms the perturbed question sequence $Q^{\text{RD}}$ with $n+\delta^{\text{RD}}$ elements. In $Q^{\text{RD}}$, all original questions from $Q$ are preserved in the original order with potential intervals, and each selected question in $X$ appears one additional time at random positions. Perturbation RD can be regarded as element addition of $Q$.
\end{itemize}

Real-world multi-turn dialogues between users and dialogue systems are inherently noisy, and often contain topic shifts, repetitions, or contextual omissions\cite{mo2025surveyconvsearch, shi2023large}, which requires dialogue systems to possess strong robustness. Perturbation RS, RR and RD are three fundamental dialogue-level perturbations. They simulate complicated real-world usage of dialogue systems in diverse situations where users may pose questions in shifted, incomplete, or redundant contexts, forming the erratic conversational scenarios\cite{traum1995computational}. Specifically, the round shuffle (RS) perturbation models random conversational and abrupt topic shifting \cite{ConditionalQuestions, hwang-etal-2024-mp2d}. The round reduction (RR) perturbation simulates context omission or the user's implicit assumption of prior knowledge\cite{clark1991grounding,pan2019improving}. The round duplication (RD) perturbation reflects redundant querying or repeating for confirmation\cite{pickering2004toward,jang-etal-2024-itercqr}. It is necessary for testing approaches to simulate such noisy scenarios to stress test LLM-based dialogue systems, and failing to properly handle such scenarios reveals the incapability of the dialogue system. To further construct varied follow-up test cases, four derivative perturbations are employed, they are:

\begin{itemize}
    \item \textbf{Perturbation RSR}: round shuffle and reduction:
    \begin{align}
        Q^{\text{RSR}} = P^\text{RSR}(Q) = P^\text{RR}(P^\text{RS}(Q)).
        \label{eq:p4rsr}
    \end{align}
    Perturbation RSR is a combined operation of RS and RR. It shuffles the original question sequence then deletes $\delta^{\text{RR}}$ randomly chosen questions to form a perturbed question sequence. 
    
    \item \textbf{Perturbation RSD}: round shuffle and duplication:
    \begin{align}
        Q^{\text{RSD}} = P^\text{RSD}(Q) = P^\text{RD}(P^\text{RS}(Q)).
        \label{eq:p5rsd}
    \end{align}
    Perturbation RSD is a combined operation of RS and RD, which shuffles then duplicates $\delta^{\text{RD}}$ randomly chosen questions, inserts the chosen questions into random positions to form a perturbed question sequence. 

    \item \textbf{Perturbation RRD}: round reduction and duplication:
    \begin{align}
        Q^{\text{RRD}} = P^\text{RRD}(Q) = P^\text{RD}(P^\text{RR}(Q)).
        \label{eq:p6rrd}
    \end{align}
    Perturbation RRD is a combined operation of RR and RD, which deletes $\delta^{\text{RR}}$ randomly chosen questions then duplicates $\delta^{\text{RD}}$ randomly chosen questions from the remaining questions. 

    \item \textbf{Perturbation RSRD}: round shuffle, reduction and duplication:
    \begin{align}
        Q^{\text{RSRD}} = P^\text{RSRD}(Q) =  P^\text{RD}(P^\text{RR}(P^\text{RS}(Q))).
        \label{eq:p7rsrd}
    \end{align}
    Perturbation RSRD is a combined operation of all three fundamental perturbations, which shuffles first, then deletes $\delta^{\text{RR}}$, and finally duplicates $\delta^{\text{RD}}$ randomly chosen questions in the original question sequence. 
    
\end{itemize}

To sum up, those dialogue perturbations simulate more complicated information-oriented real-world requests in dialogue, and test the performance of dialogue systems. They require dialogue systems to tell if sufficient information is given in context and perform different behaviours. These perturbations further extend the test coverage of dialogue systems with rarely seen situations. The examples of perturbed multi-turn test input sequences are provided in supplementary material.

\subsection{MRs in MORTAR}
\label{sec:mr-matching}
For one specific question in the original question sequence, the context can be changed using dialogue-level perturbations. According to Eq.\ref{eq:root_mr}, if the perturbed context remains equivalent to the original context, i.e., the target question in perturbed context is asking the same fact as in the original context, the dialogue system is expected to respond with a semantically similar answer, regardless of whether the question is asked in the RBT or MT. Otherwise, for the same question with inequivalent contexts, the dialogue system is expected to respond with semantically different answers, as the questions are targeting different facts.

The equivalent context check $EC(\cdot)$ is a function that verifies whether the context of a question in a perturbed test case is equivalent to the context of the identical question in the original test case. The detailed implementation of $EC(\cdot)$ is introduced in Section \ref{sec:ecc}. Given the target question $q^r_i, i\in\{1,\dots,n^r\}$ and its context $Q^r_{i-1}$ in the perturbed test case, $EC(\cdot)$ compares whether $Q^r_{i-1}$ is equivalent to context of the identical target question $q_j$ in original test case $Q$, which is $Q_{j-1}$. If equivalent (noted as $Q^r_{i-1} \equiv Q_{j-1}$), $EC(\cdot)$ will give \texttt{True}, otherwise \texttt{False}, such that we have the following:
\begin{align}
\begin{split}
EC(Q^r_{i-1}, q^r_i) = 
\begin{cases} 
    \text{True}, & \text{if }Q^r_{i-1} \equiv Q_{j-1} \\  
    \text{False}, & \text{Otherwise}
\end{cases}
\end{split}
\label{eq:ac}
\end{align}

Given the test seed $Q$ and a set of perturbation operations $R$, each $r$ in $R$ is one of the independently executed perturbations among RS, RR, RD, RSR, RSD, RRD, and RSRD, even the same perturbation in two executions may yield different test cases. For the original test input from $Q_j$ and perturbed test input from $Q^r_i$, the dialogue system $DS(\cdot)$ respectively generates responses $o_j$ and $o^{r}_{i}$:
\begin{align}
    DS(Q_{j-1}, q_j) &\rightarrow o_j,\\
    DS(Q^{r}_{i-1}, q^r_i) &\rightarrow o^{r}_{i},
\end{align}
where $i\in\{1,\dots,n^r\}$, $j\in\{1,\dots,n\}$, $q^r_i = q_j$, $q^r_i$ is the target question in perturbed test case, $q_j$ is the identical question in original test case, $o^{r}_{i}$ and $o_j$ are outputs in MT and RBT respectively, and $Q^{r}_{i-1}$ and $Q_{j-1}$ are contexts in MT and RBT respectively. 

To measure the semantic similarity of two utterances, we use the natural language embedding model ($\text{Emb}(\cdot)$) to calculate the embedding vector of each utterance, then calculate the cosine similarity of the two embedding vectors. If the cosine similarity reaches the threshold $\epsilon$, the two utterances will be regarded as semantically similar, otherwise semantically different.

Using equivalent context check function, MR1 and MR2 are defined as follows:
\begin{itemize}
\item \textbf{MR 1: Context-Preserving MR.} 
If a perturbed context $Q^{r}_{i-1}$ is equivalent to $Q_{j-1}$ in terms of supplementary information to the target question:
\begin{align}
    EC(Q^r_{i-1}, q^r_i) = EC(Q_{j-1}, q_j) = \text{True},
\label{eq:mr1}
\end{align}
the dialogue system is expected to produce an answer semantically similar to the original expected response:
\begin{align}
\text{Expect: } \text{cos}(\text{Emb}(o^r_i), \text{Emb}(a_j)) \geq \epsilon.
\label{mr1}
\end{align}

\item \textbf{MR 2: Context-Altering MR.} 
If perturbed context $Q^r_{i-1}$ of $q^r_i$ lacks critical information to clarify $q^r_i$ or is inequivalent to $Q_{j-1}$:
\begin{align}
    EC(Q^r_{i-1}, q^r_i) \neq EC(Q_{j-1}, q_j),
\label{eq:mr2}
\end{align}
the dialogue system is expected to produce an answer semantically different from the original expected response:
\begin{align}
\text{Expect: } \text{cos}(\text{Emb}(o^r_i), \text{Emb}(a_j)) < \epsilon.
\label{mr2}
\end{align}
\end{itemize}

All perturbations can be fitted with MR1 or MR2. Given the context check results, it can be determined which MR and perturbation shall be matched for bug detection. In addition to individual MRs, we further propose group-based MRs that expect dialogue systems to adapt to and ensure consistent behaviour among different versions of context-changed questions. 

For an original target question $ q_j $ from the original dataset, assuming there are multiple perturbed question sequences where the identical target question appears at different rounds with different contexts. We name the same question under different contexts as different \textit{versions} of the original question $q_j$. After executing different perturbations in $R$, all different versions of $q_j$ form a group $q^R_j$, and all outputs of the dialogue system under test form a group $o^R_j$:
\begin{align}
    q^R_j &= \{q^r_i | q^r_i = q_j, r \in R\}\\
    o^R_j &= \{o^r_i | o^{r}_{i} = DS(Q^{r}_{i-1}, q^r_i), q^r_i = q_j, r \in R\}
\end{align}

The context of each $q^r_i$ in corresponding test case may be equivalent or inequivalent to the original context of $q_j$ in test seed. We split the outputs $o^R_j$ into the output group of context-preserving perturbations of the original question, $o^{R+}_j$, and the output group of context altered perturbations, $o^{R-}_j$:
\begin{align}
    o^{R+}_j &= \{o^r_i|EC(Q^r_{ i-1}, q^r_i)=\text{True}, q^r_i \in q^R_j\}, \\
    o^{R-}_j &= \{o^r_i|EC(Q^r_{ i-1}, q^r_i)=\text{False}, q^r_i \in q^R_j\} .
\end{align}

\begin{itemize}    
    \item \textbf{MR 3: Inner-Group Consistency MR.} 
    Within group $o^{R+}_j$, different versions of $q_j$ have equivalent contexts to the original, the dialogue system should not produce highly divergent answers, meaning the maximum difference within the group should be below threshold $\epsilon$:
    \begin{align}
    \text{Expect:}\!\min_{o_1, o_2 \in o^{R+}_j}\!\text{cos}(\text{Emb}(o_1),\! \text{Emb}(o_2))\! \geq\! \epsilon.
    \label{eq:mr3}
    \end{align}
    \item \textbf{MR4: Inter-Group Divergence MR.} 
    For two versions of $q_j$ from the two groups, the system’s outputs must be different and greater than $\epsilon$:
    \begin{align}
    \text{Expect:}\!\max_{o_1\in o^{R+}_j,o_2\in o^{R-}_j}\!\text{cos}(\text{Emb}(o_1),\! \text{Emb}(o_2))\! <\! \epsilon.
    \label{eq:mr4}
    \end{align} 
\end{itemize}

\begin{figure}[]
  \centering
  \includegraphics[width=\linewidth]{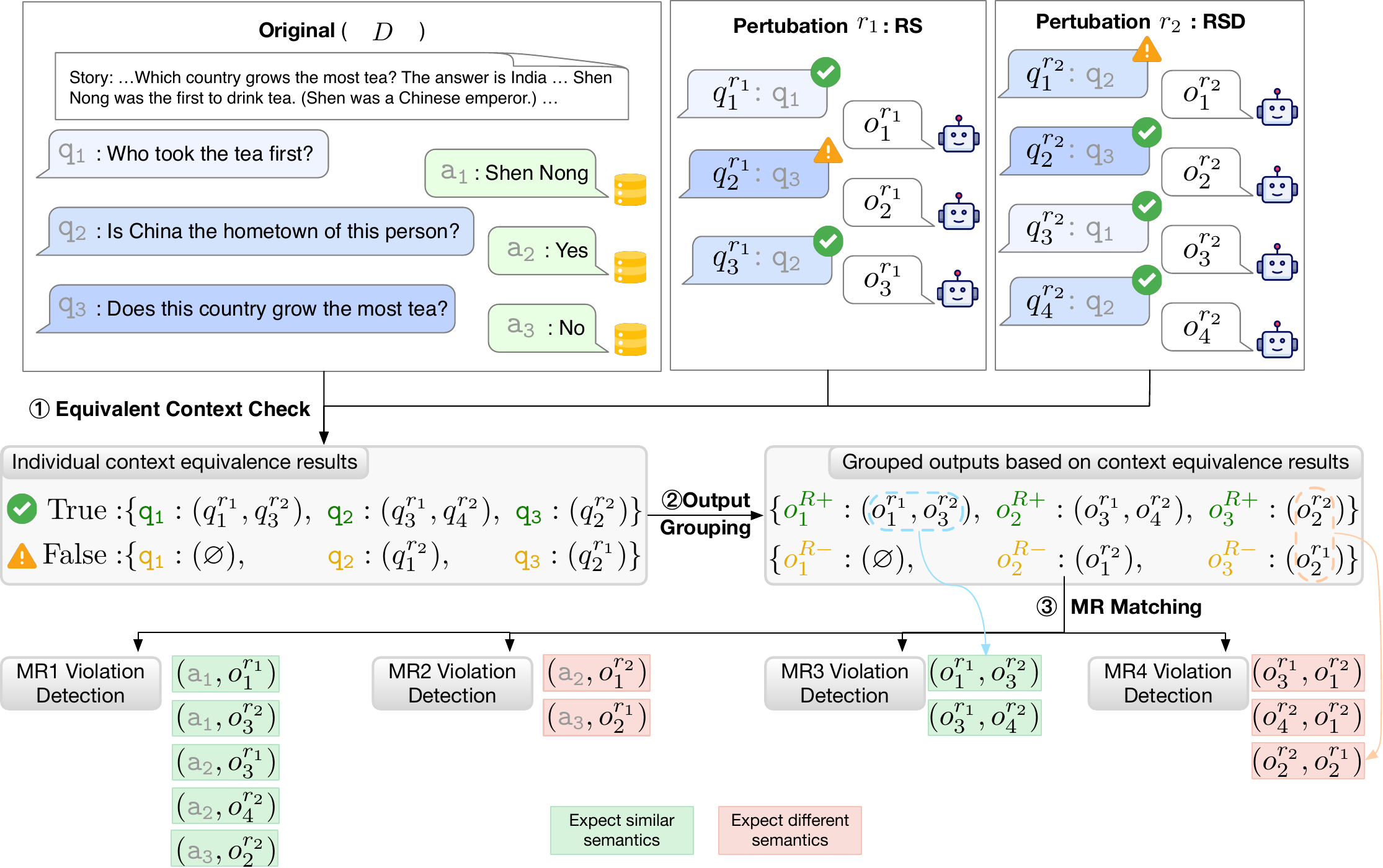}
  \caption{Perturbation and MR violation detection. Given the perturbed test inputs and original test case, using the equivalent context check results, the outputs are matched with corresponding MR violation detection.}
  \label{fig:MR_Conflict}
\end{figure}

As shown in Fig.~\ref{fig:MR_Conflict}, given the original multi-turn dialogue ($D=[(q_1,a_1),(q_2,a_2),(q_3,a_3)]$) as a test seed, all questions can be regrouped as $Q$: $Q=[q_1, q_2, q_3]$. Two perturbations $r_1$ (RS) and $r_2$ (RSD) are adopted and generates two new question sequences: $Q^{r_1}=[q^{r_1}_1,q^{r_1}_2,q^{r_1}_3]=[q_1,q_3,q_2]$ and $Q^{r_2}=[q^{r_2}_1,q^{r_2}_2,q^{r_2}_3,q^{r_2}_4]=[q_2,q_3,q_1,q_2]$ respectively. The questions in $Q$ may appear at different rounds in perturbed test cases, such that the identical question may have varied context and targeting information in perturbed test case when compared with the original test case. For example, $q_1$ is the first question in $Q$, and it is asked in the first round in $Q^{r_1}$ as $q^{r_1}_1$ and the third round in $Q^{r_2}$ as $q^{r_2}_3$.

The variance results in different contexts equivalence, some questions' contexts remain equivalent to the identical question in the original contexts while the others do not. For example, the context of $q^{r_2}_3$ in $Q^{r_2}$ is $Q^{r_2}_{2}=[q^{r_2}_{1},q^{r_2}_{2}]=[q_2,q_3]$, the identical question of $q^{r_2}_3$ in original test case $Q$ is $q_1$, and the context of $q_1$ is $Q_0$ which is \texttt{None}. As $q_1$ is an independently answerable question without reliance on any context questions, the context of $q^{r_2}_3$ can be regarded as equivalent to the context of $q_1$, i.e., $EC(Q^{r_2}_2,q^{r_2}_3)=\text{True}$. Meanwhile, $q_3$ is the third question in $Q$, and it is asked in the second round in $Q^{r_1}$ as $q^{r_1}_2$. Different from the aforementioned example, $EC(Q^{r_1}_1, q^{r_1}_2)=\text{False}$, where $Q^{r_1}_1=[q^{r_1}_1]=[q_1]$. Similarly for all questions in perturbed test cases, the equivalent context check (Step \ding{172} in Fig.~\ref{fig:MR_Conflict}) is conducted to obtain the individual context equivalence results. For example, $q_2$ is the second question in $Q$, it is also asked in the third round in $Q^{r_1}$ as $q^{r_1}_3$, and in both the first and fourth round in $Q^{r_2}$ as $q^{r_2}_1$ and $q^{r_2}_4$ respectively. The individual context equivalence results as following: $EC(Q^{r_1}_2,q^{r_1}_3)=\text{True}$, $EC(Q^{r_2}_0,q^{r_2}_1)=\text{False}$, and $EC(Q^{r_2}_3,q^{r_2}_4)=\text{True}$. The results are also presented in group for simplicity: $\text{True}: \{q_2:(q^{r_1}_3, q^{r_2}_4)\}$ and $\text{False}:\{q_2:(q^{r_2}_1)\}$. 

Based on the individual context equivalence results, we group the outputs of system under test based on the questions' context equivalence to each corresponding question in $Q$ (Step \ding{173} in Fig.~\ref{fig:MR_Conflict}). For example, $q_1$ is identical to $q^{r_1}_1$ and $q^{r_2}_3$, and $EC(Q^{r_1}_0, q^{r_1}_1)=EC(Q^{r_2}_2, q^{r_2}_3)=\text{True}$, such that the corresponding outputs $o^{r_1}_1$ and $o^{r_2}_3$ are grouped in $o^{R+}_1$, where $R=\{r_1, r_2\}$.

In step \ding{174}, using the grouped context equivalence results, MRs are adaptively matched and used to detect bugs. For example, MR1 is an individual MR and $o^{r_2}_3$ is an element of $o^{R+}_1$, according to the definition of MR1 in Eq.\ref{eq:mr1}, we expect $a_1$ and $o^{r_2}_3$ to be semantically similar, as $q_1$ and $q^{r_2}_3$ are asking the identical information, although they have varied contexts. If $a_1$ and $o^{r_2}_3$ are semantically different, MR1 will be violated, indicating a bug is revealed in the dialogue system under test. Similarly for MR2, $o^{r_1}_2$ is an element of $o^{R-}_3$, according to the definition of MR2 in Eq.\ref{eq:mr2}, we expect $a_3$ and $o^{r_1}_2$ to be semantically different. If $a_3$ and $o^{r_1}_2$ are semantically similar, MR2 will be violated and a bug will be revealed in the dialogue system under test. Violation to MR2 indicates the dialogue system not performing well in adapting to the context in multi-turn dialogues. For group-based MRs, take $o^{R+}_1$ for example, two elements belong to this group: $o^{R+}_1:\{o^{r_1}_1,o^{r_2}_3\}$, according to the definition of MR3, we expect $o^{r_1}_1$ and $o^{r_2}_3$ to be semantically similar. And for MR4, $o^{R+}_3$ has an element $o^{r_2}_2$ and $o^{R-}_3$ has an element $o^{r_1}_2$, such that we expect $o^{r_2}_2$ and $o^{r_1}_2$ to be semantically different, as their questions are targeting different information in inequivalent contexts.

To our knowledge, the above formalisation is the first attempt to model the multi-turn MT for dialogue systems. It is worth noting that MORTAR's process is different from single-turn metamorphic testing methods. In existing methods, the perturbation or test seed and MR violation detection are coupled, such that the testing is inevitably semi-automated as manual selection or predefining the MR for violation detection is necessary. In multi-turn MT, the effects of dialogue-level perturbations are determined later with the equivalent context check process. This mechanism allows MORTAR to enhance the diversity of test cases and maximise the usage of test seeds. Given that $EC(\cdot)$ is automated, the multi-turn metamorphic testing can be fully automated. 

\subsection{Equivalent Context Check}
\label{sec:ecc}

\begin{figure}[!t]
  \centering
  \includegraphics[width=1\linewidth]{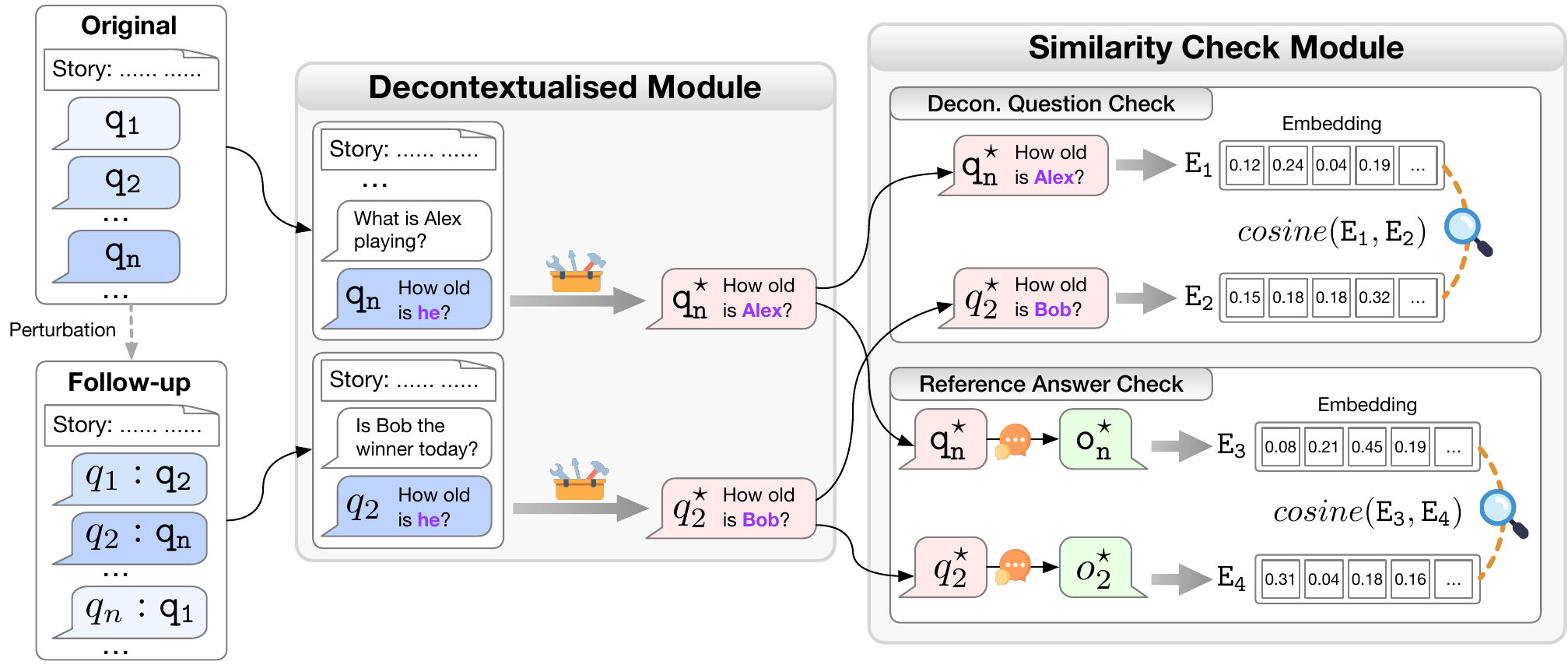}
  \caption{Equivalent context check pipeline in MORTAR. }
  \label{fig:ECC_pipelines}
\end{figure}

In Equation~\eqref{eq:ac}, the equivalent context check ($EC(\cdot)$) gives whether the context of a question in a perturbed test case is equivalent to the context in the original test case, and it is also a key component to judge whether an MR is applicable or not. As the effect of dialogue-level perturbations is dynamic, the answer expectation of some questions in perturbed test cases may differ from answers of the original question or the same question in other perturbations, such that different MRs shall be adopted according to the actual situation. To handle this issue and implement automated metamorphic testing, it is necessary to use the equivalent context check and dynamically match the appropriate MR according to the context of each question in the perturbed test case. To the best of our knowledge, there is no available approach in the literature that formalise or automate such a notion of contextual equivalence. As shown in Fig.\ref{fig:ECC_pipelines}, we propose an effective design of equivalent context check composed of 1) the decontextualisation, and 2) the similarity check.

\subsubsection{Decontextualisation}
Inspired by the question rewrite task in natural language processing~\cite{elgohary2019can}, which extracts information from dialogue context and rewrite the target question into an independently answerable question, MORTAR adopts a prompt-based approach and use an LLM to rewrite and decontextualise the target questions with perturbed context. The prompt template is provided in supplementary material. 

Given the context information, i.e., the reference article and the question sequence from context, a target question can be rewritten as a context-independent question that carries sufficient context on its own, while asking the same fact as the context-dependent form. For example, in Fig.\ref{fig:ECC_pipelines}, the target question ``How old is he'' is $\text{q}_n$ in original test case and the $q_2$ in perturbed test case, and the perturbation can be RS or other derivative perturbations (see Section \ref{sec:dialogue-level-perturbations}). This target question cannot be answered independently unless the ellipsis information (who is ``he'') is clarified in the context. 

In the original test case, the target question is decontextualised as ``How old is Alex?''. However, in the perturbed test case, it is decontextualised as ``How old is Bob?'' as the context of perturbed test case is \textbf{inequivalent} to the context of the original test case. Pronouns are common anaphoric expressions\cite{sukthanker2020anaphora} in multi-turn dialogues that cause context reliance of target questions, while MORTAR is also capable of handling various types of ellipsis\cite{zhang2020ellipsis}. For example, verbal ellipsis involves the omission of a verb phrase that appeared earlier in the dialogue. If the context contains ''Alex can speak French,'' and the target question is ''Can Bob?'', MORTAR identifies the missing predicate and rewrites the question as ''Can Bob speak French?''. If the contexts of the target question in test seed and the perturbed test case are equivalent, the decontextualised question will be similar in semantics and asking the same fact. Otherwise, the inequivalent contexts will result in different decontextualised questions, and the similarity check step will detect the difference to give equivalent context check result.

\subsubsection{Similarity Check}
The similarity check step compares whether two decontextualised questions are semantically similar, and the similarity indicates the equivalence of corresponding contexts. In MORTAR, similarity check is composed of two parts: 1) the decontextualised question check, and 2) the reference answer check. 

In the question similarity check, the natural language embedding model is used to calculate the semantic embedding vectors of the decontextualised questions ($\mathtt{E_1}=\text{Emb}(\mathtt{q^*_n})$, $\mathtt{E_2}=\text{Emb}(q^*_2)$). The similarity of the decontextualised questions is given by the cosine similarity of the two embedding vectors: $Sim_Q=\text{cos}(\mathtt{E_1}, \mathtt{E_2})$.

In practice, the question decontextualisation occasionally rewrites questions with too much or irrelevant information. For example, the question ``Where was he coming from?'' is rewritten as ``From where was Guy coming?'' in the original test case, and ``From where was Guy, who proposed to Virginia Duge, coming?'' in the perturbed test case, since the perturbed test case has an additional round in context: ``who proposed to Virginia Duge''. Although they can be answered with the same answer, the similarity of the two decontextualised questions is not as high as other question pairs. Hence, merely relying on the question similarity check could mistakenly output ``False'' for such cases and result in increased false positive rates. To handle this issue, we add a reference answer check process. We use a prompt-based approach to obtain the reference answers of the two decontextualised questions ($\mathtt{o^*_n}$ and $o^*_2$) with single-turn input, and compute the similarity of the embeddings ($\mathtt{E_3}=\text{Emb}(\mathtt{o^*_n})$ and $\mathtt{E_4}=\text{Emb}(o^*_2)$) of the reference answers of two decontextualised questions. The reference answers check does not generate test oracles, but used to filter out the noise in the output of question decontextualisation. 

In this way, the overall similarity score is calculated by averaging the similarity of the decontextualised questions and the similarity of the reference answers. If the overall similarity score reaches a preset threshold $\epsilon_{EC}$, the equivalent context check will indicate ``True'', otherwise ``False'':

\begin{align}
\begin{split}
EC(\cdot) = 
\begin{cases} 
    \text{True}, & \text{if } \frac{\text{cos}(\mathtt{E_1}, \mathtt{E_2}) + \text{cos}(\mathtt{E_3}, \mathtt{E_4})}{2} \geq \epsilon_{EC} \\  
    \text{False}, & \text{Otherwise}
\end{cases}
\end{split}
\label{eq:ecc_score}
\end{align}

To summarise, different questions in a perturbed test case may be matched with different MRs for violation detection, and this matching is driven by the automated equivalent context check ($EC(\cdot)$). Instead of delegating the equivalent/inequivalent verdict to an end-to-end LLM prompt, $EC(\cdot)$ in MORTAR limits the usage of LLM to constrained natural language processing tasks, and each step of $EC(\cdot)$ can therefore be independently verified, as reported in Section~\ref{sec:results}.

\section{Design of Experiments}
\label{sec:exp}
In this section, we introduce the research questions, setting of experiments, baseline methods, the dataset, dialogue systems under test, and relevant implementation details.

\subsection{Research Questions}
\label{subsec:RQ}

\textit{\textbf{RQ 1:}} How does MORTAR perform in detecting bugs in LLM-based dialogue systems? We answer this research question by comparing the effectiveness and robustness of MORTAR with the baseline methods.

\textit{\textbf{RQ 2:}} What is the quality of the bugs revealed by MORTAR? In this RQ, we analyse the characteristics and quality of detected bugs in terms of diversity, precision and uniqueness.

\textit{\textbf{RQ 3:}} How effective are the components of MORTAR? In this RQ, we conduct an ablation study and investigate the contribution of the seven perturbations and four MRs to the overall performance, and the effectiveness of equivalent context check component.

\textit{\textbf{RQ 4:}} How sensitive is MORTAR to its hyperparameter configurations? In this RQ, we examine the performance of MORTAR under different perturbation rate and semantic similarity threshold.

\subsection{Data Preparation}
Regarding the test dataset, we have the following expectations: \textit{a}) focusing on fundamental capabilities of testing dialogue systems; \textit{b}) have two human participants with one posing questions and another answering the questions; \textit{c}) questions might have reliance on dialogue history; \textit{d}) questions can be independently answered when given sufficient information so that the question can be decontextualised. CoQA~\cite{reddy2019coqa} fundamentally meets all the above requirements. CoQA is a multi-turn reading comprehension dataset with topic in multiple domains. Each record is composed of a story and a multi-turn dialogue record. The multi-turn dialogue is produced by a human question-asker and a human answerer. It comprises a training set and a development set. We adopt the development set as test seeds. There are a total of 500 dialogues with 7983 questions. 

\subsection{Baselines}
To the best of our knowledge, currently there are no multi-turn dialogue testing methods that provide a compatible and judge-free test oracle with explicit pass/fail criteria on fixed seed dialogues, hence no existing multi-turn baseline can be directly used for a fair performance comparison with MORTAR. Popular multi-turn evaluation benchmarks that rely on LLM judges yield graded or preference scores rather than MR-grounded pass/fail outcomes, and therefore do not constitute testing oracles. In contrast, single-turn metamorphic testing methods define automated, label-free oracles and are the closest feasible comparators. Therefore, we adapt two representative single-turn MT baselines: METAL\cite{hyun2024metal} and QAAsker\cite{chen2021qaasker}. METAL is one of the latest and most effective MT frameworks for single-turn LLM testing. We use the four most effective single-turn MRs in METAL: the model’s outputs on the original and the perturbed input should not differ, the perturbations are \textit{synonym-replacement}, \textit{add random word}, \textit{introduce-typos}, and \textit{convert-to-leet-format}. QAAsker is an effective and representative single-turn metamorphic testing method for QA systems. It constructs follow-up questions using three MRs with full test automation. 

As the effect of single-turn MT may change the information content in each turn, we adopt their original implementation of perturbations on each input question and concatenate the perturbed question with original context to generate the follow-up test cases. This unified adaptation of single-turn metamorphic testing in multi-turn testing ensures the original single-turn MRs still strongly hold and the context remains fully informative to each perturbed question, such that they function as comparable baselines to MORTAR.

\subsection{Test Objects}
\label{sec:test_objects}
We select six open-source conversational LLMs and two representative closed-source LLMs and prompt each of them to function as a dialogue system (DS). In the prompts, the story is first provided as a reference, then the LLMs are required to answer questions with \textit{concise and short} answer to reduce unrelated information, and they are required to answer ``unknown'' if they do not know the answer. When testing these dialogue systems, we simulate a user having multi-turn dialogues with the dialogue system. We feed the dialogue system with the question sequence in each metamorphic test case along with the dialogue history, then collect the generated answer to analyse response quality. 

\begin{table}[]
\caption{The dialogue systems under test}
\label{tab:dsuts}
\centering
\adjustbox{max width=\linewidth}{
\begin{tabular}{llll}
\toprule
\textbf{} & \textbf{Model} & \textbf{Type} & \textbf{Model Size} \\
\midrule
DS1 & Qwen2-0.5B-Instruct~\cite{qwen2} & Open-source & 494M\\
DS2 & Qwen2-1.5B-Instruct~\cite{qwen2} & Open-source & 1.54B\\
DS3 & Qwen2-7B-Instruct~\cite{qwen2} & Open-source & 7.62B\\
DS4 & Mistral-7B-Instruct-v0.3~\cite{mistral7b} & Open-source & 7.25B \\
DS5 & Llama-3-8B-Instruct~\cite{llama3modelcard} & Open-source & 8.03B\\
DS6 & Gemma-2-9b-it~\cite{gemma_2024} & Open-source & 9.24B\\
DS7 & GPT-5-nano~\cite{openai_gpt5nano_2025} & Closed-source (Commercial) & Unknown\\
DS8 & GPT-5~\cite{openai_gpt5_2025} & Closed-source (Commercial) & Unknown\\
\bottomrule
\end{tabular}}
\end{table}

The selected LLMs for each dialogue system are presented in Table~\ref{tab:dsuts}. The open source models used are all instruction finetuned and their sizes range from 0.5 to 9 billion parameters, and the inference temperature is setup to be 0.2 and top-p parameter is set up to be 0.9 to reduce hallucination and repetitive degeneration while maintain relative deterministic behaviour in multi-turn dialogue\cite{holtzman2019curious,ji2023survey}. These dialogue systems can be regrouped into three sets: the first group is composed of DS1, DS2, and DS3. They are the same sourced models with different parameter sizes. The second group is composed of DS3, DS4, DS5 and DS6. They are all based on popular and recently published language models with similar model sizes. The third group is composed of DS7 and DS8 and they are based on representative closed-source LLMs GPT-5-nano and GPT-5. GPT-5 is one of the latest closed-source LLMs. GPT-5-nano is a fast and cost-effective version of GPT-5 and is more suitable for latency-sensitive environments where the quality of output shall not be compromised. In our experiment, the testing for open-source LLM-based dialogue systems is conducted locally. Using the instruction finetuned model weights, we implemented a series of dialogue systems powered by conversational LLMs with different sources and model sizes. The testing of closed-source models is conducted with API calls. We set the reasoning effort and output verbosity to low to ensure more deterministic behaviour and concise response. 

\subsection{Criteria tailored to multi-turn metamorphic testing}
\label{sec:criteria}

\begin{table}[]
    \centering
    \caption{Metrics used to evaluate the performance of dialogue systems in multi-turn testing.}
    \adjustbox{max width=\linewidth}{%
    \begin{tabular}{l|p{8cm}}
        \toprule
       Metrics & Explanation \\
        \midrule

        Bugs & The number of bugs, a bug is an incorrect answer when compared with the ground truth answer or a violation with MR. \\

        \midrule

        Bugs per test case (B/TC) & The number of bugs divided by the number of test cases. \\

       \midrule

       Turn failure rate (Rate\textsuperscript{+}) & The number of detected bugs divided by the number of tested turns. \\

       \midrule
       
       L1-Bugs & The number of L1-Bugs, an L1-Bug is revealed by the same question in both RBT and MT. \\

       \midrule
       
       L2-Bugs & The number of L2-Bugs, an L2-Bug is revealed in MT using the correctly answered question from partially passed test case in RBT. \\

       \midrule
       
       L3-Bugs & The number of L3-Bugs, an L3-Bug is revealed in MT using any question from fully passed test cases in RBT. \\

        \midrule
    \end{tabular}}
    \label{tab: Test criteria summary}
\end{table}

When testing the dialogue system with multi-turn test cases, we use a series of multi-turn specific metrics to quantitatively describe the performance of a testing method, the metrics are summarised in Table~\ref{tab: Test criteria summary}. They are adapted from existing single-turn testing metrics and extended to multi-turn testing to provide fine-grained quantitative performance measure of dialogue systems. Following the established naming convention in MT\cite{shen2022qaqa,wang2024kgit}, a \textit{bug} is defined as a mistakenly answered question in RBT or a violation to MR in MT of the system under test, and a \textit{detection} is the identification of those mistakes in a piece of output. A multi-turn test case is capable of revealing multi-turn bugs through verifying the output of each turn. If all questions are correctly answered and no MR is violated, the test case will be regarded as a \textit{passed test case}. Otherwise, if one or more outputs are wrong or conflict the MR, the test case will be regarded as a \textit{failed test case} (FTC), the turn with a bug will be regarded as a \textit{failed turn}. The number of bugs per test case measures the density of bugs. The turn failure rate (Rate\textsuperscript{+}) is a general metric in testing, measuring the overall performance of dialogue systems and the efficiency of testing methods. In RBT, it indicates the ratio of questions that are not responded with expected answers. In metamorphic testing, it measures the ratio of MR violations, given by:
\begin{align}
\text{Rate}^\text{+} = \frac{\text{Total number of detected bugs}}{\text{Total number of detections}} \times 100\%. \notag
\end{align}

As defined in Section\ref{sec:MTMT}, a bug revealed using RBT is a question that the system under test responds with different answer from the test oracle, while a bug revealed using MT is a pair of outputs that violates MR. To delve into the type of revealed bugs, we categorise bugs based on how they are exposed when compared with RBT using the test seed. 

An \textbf{L1-Bug} is a bug revealed by both the RBT and MT, i.e., a question is answered incorrectly in both RBT and MT. An \textbf{L2-Bug} is revealed when the RBT already reveals one or more bugs using the test seed, while other correctly answered questions are incorrectly answered in MT. An L2-Bug is a new round-level bug. An \textbf{L3-Bug} is a bug revealed in the condition that, all questions from a test seed are answered correctly in RBT, while at least one question is incorrectly answered by the system under test in MT. An L3-Bug is a new dialogue-level bug, and can be regarded as a successful reuse of a fully-passed test seed from RBT. 

Take $q_3$ in Fig.\ref{fig:MR_Conflict} for example, $q_3$ is the third question in RBT using the test seed and the second question ($q^{r_2}_2$) with perturbation $r_2$ in MT. If $o^{r_2}_2$ and $a_3$ violates MR1 ($EC(Q^{{r_2}}_1, q^{r_2}_2)=True$), it will be categorised as an 1) L1-Bug if: $q_3$ is incorrectly answered in RBT, or 2) L2-Bug if: $q_3$ is correctly answered in RBT while at least one other questions (can be $q_1$ or $q_2$) are incorrectly answered, or 3) L3-Bug if: all other questions in this test seed ($q_1$, $q_2$, and $q_3$) are all correctly answered in RBT.

The manual check will use F1 Score to assess the subjects, given by:
\begin{align}
\text{F1} = \frac{2\times \text{Precision}\times\text{Recall}}{\text{Precision}+\text{Recall}} \notag
\end{align}

where recall is given by:
\begin{align}
\text{Recall} = \frac{\text{True positive}}{\text{True positive}+\text{False negative}} \notag
\end{align}

and the precision is given by:
\begin{align}
\text{Precision} = \frac{\text{True positive}}{\text{True positive}+\text{False positive}}. \notag
\end{align}

Additionally, we use coefficient of variation (CV) to measure the spareness, calculated as:
\begin{align}
\text{CV} &= \frac{\sigma}{\mu}, \notag
\end{align}
where $\sigma$ and $\mu$ are the standard deviation and mean.

\subsection{Antecedent testing: RBT with original dataset}
\label{sec:RBT}
\subsubsection{Result of RBT}
\begin{table}
    \centering
    \caption{Test result of RBT using original test cases.}
    \adjustbox{max width=\linewidth}{%
    \begin{tabular}{l|rrrrrrrr}
        \toprule
        & DS1 & DS2 & DS3 & DS4 & DS5 & DS6 & DS7 & DS8 \\
        \midrule
        Failed test cases & 492 & 446 & 369 & 460 & 292 & 279 & 401 & 271 \\
        Ratio of failed test cases & 98.40\% & 89.20\% & 73.80\% & 92.00\% & 58.40\% & 55.80\% & 80.20\% & 54.20\% \\
        Bugs & 3,386 & 1,965 & 1,147 & 1,589 & 524 & 469 & 1,153 & 462 \\
        Rate\textsuperscript{+} & 42.42\% & 24.61\% & 14.37\% & 19.90\% & 6.56\% & 5.87\% & 14.44\% & 5.79\% \\
        \bottomrule
    \end{tabular}}
    \label{tab:RBT_result}
\end{table}

When conducting RBT on all dialogue systems under test with original test cases, all dialogue systems are exposed to the same test set of 7,983 questions from 500 original dialogue test cases. According to the result in Table~\ref{tab:RBT_result}, substantial unexpected behaviour is observed. Among all dialogue systems, DS8 performs the best. The turn failure rate of DS8 is 86\% lower than DS1. We generate follow-up test cases to further test the dialogue systems.

\subsubsection{Justification of answer similarity threshold ($\epsilon$) in multi-turn testing scenarios}
\label{rbt_epsilon_mt}

\begin{table}[]
    \centering
    \caption{The F1 scores of different multi-turn answer similarity threshold. Higher is better. $\epsilon=0.6$ reaches the highest F1 score.} 
    \adjustbox{max width=\linewidth}{%
    \begin{tabular}{l|rrrrrr}
    \toprule
        $\epsilon$  & 0.4 & 0.5 & 0.6 & 0.7 & 0.8 & 0.9  \\
        \midrule
        F1 score & 0.841  & 0.867 &\textbf{0.874} & 0.868 & 0.788 & 0.643  \\
         \bottomrule
    \end{tabular}}
    \label{tab:RBT_epsilon}
\end{table}

We conduct a manual check on RBT test results to justify the answer similarity threshold for multi-turn scenarios. The manual labelling is conducted by two authors. We randomly sample 20 responses of each dialogue system in RBT to form a total of 160 evaluated responses, and two authors manually label if the response is correct or not and give binary judgement. After independent labelling, the two authors discuss on the diverged judgements and create consensus labels. The manual labels are then used to verify the F1 score under different values of $\epsilon$. According to the Table~\ref{tab:RBT_epsilon}, $\epsilon=0.6$ results in the highest F1 score, indicating best agreement with human judgements.

\subsubsection{Generation of follow-up test cases}

\begin{table}[]
    \centering
    \caption{The summary of follow-up test dataset} 
    \adjustbox{max width=\linewidth}{%
    \begin{tabular}{l|rrrr}
    \toprule
        & QAAsker & METAL & MORTAR \\
        \midrule
       Follow-up test cases & 1500 & 2,000 & 3,500 \\
       Maximum detectable bugs & 9,796 & 24,803 & 57,010 \\
       \bottomrule
    \end{tabular}}
    \label{tab:data_gen}
\end{table}

We use the original test dataset to generate the perturbed test cases with MORTAR and baseline methods to further test the dialogue systems. The follow-up testing datasets are summarised in Table~\ref{tab:data_gen}. Given the original dataset, all testing methods can generate enlarged follow-up test dataset. It is worth noting that the MORTAR-generated test cases cannot be used in RBT, as they only carry verifiable metamorphic relations rather than ground truth answers. The number of different categories of bug (L1 to L3, defined in Section \ref{sec:criteria}) can also be interpreted as the comparison between MORTAR and RBT.

\subsection{Experiment Environment and Hyper Parameter Settings}
The local machine is equipped with 32-core processors, 64GB memory and an RTX 3090 GPU. The system runs Ubuntu 24.04 LTS and Python 3.9. Groq \cite{groq} is an LLM inference cloud service provider. The equivalent context check is implemented with the open-sourced gpt-oss-120B model accessed through Groq. This is the largest open-weight model freely accessible on Groq's production tier. We use the gte-large-en-v1.5 model \cite{zhang2024mgte} to compute the 1024-dimension embedding vectors. This embedding model achieved strong performance on English natural language embedding within the same model size category\cite{muennighoff2022mteb}. These configurations supports accurate equivalent context check and sentence embedding vector computation.

The semantic similarity threshold in equivalent context check $\epsilon_{EC}$ in Section\ref{sec:ecc} is set up to be 0.8 (justified in Section \ref{sec:RQ4}), the answer similarity threshold $\epsilon$ in Section~\ref{sec:MTMT} is set to be 0.6 (justified in Section \ref{sec:RBT}). By default, the perturbation ratios in RR, RD and derivative perturbations in Section~\ref{sec:dialogue-level-perturbations} are all set to be 30\%. We further analyse sensitivity to hyperparameters in RQ4.

\section{Results}
\label{sec:results}
In this section, we present the empirical results to answer RQs presented in Section~\ref{subsec:RQ}, and discuss the information extraction, false positive detections, and the performance evaluation of different dialogue systems.

\subsection{RQ1. Performance of MORTAR}
To answer RQ1, we evaluate the overall performance of MORTAR and compare it to the baselines in revealing dialogue system bugs with follow-up testing in terms of effectiveness and robustness. The effectiveness is measured with the performance of revealing bugs in terms of number of bugs and turn failure rate. The efficiency is measured as the number of bugs revealed per test case. The robustness is measured by the performance variation when testing different dialogue systems.

\subsubsection{Effectiveness}

\begin{table}[!t]
    \centering
    \caption{Testing the dialogue systems with MORTAR and baseline methods. B/TC: number of bugs per test case. Rate\textsuperscript{+}: turn failure rate} 
    \adjustbox{max width=\linewidth}{%
    \begin{tabular}{l|rrr|rrr|rrr}
        \toprule
        \multirow{2}{*}{DS} & \multicolumn{3}{c}{QAAsker} & \multicolumn{3}{c}{METAL} & \multicolumn{3}{c}{MORTAR} \\
        \cmidrule(lr){2-4} \cmidrule(lr){5-7} \cmidrule(lr){8-10}
            & Bugs & B/TC & Rate\textsuperscript{+} & Bugs & B/TC & Rate\textsuperscript{+} & Bugs & B/TC & Rate\textsuperscript{+} \\
        \midrule
        DS1 & 5,037 & 3.36 & 51.42\% & 12,940 & 6.47 & 52.17\% & 30,009 & \textbf{8.57} &\textbf{52.64\%} \\        
        DS2 & 3,399 & 2.27 & 34.70\% & 8,505 & 4.25 & 34.29\% & 22,166 & \textbf{6.33} & \textbf{38.88\%} \\
        DS3 & 2,558 & 1.71 & 26.11\% & 6,093 & 3.05 & 24.57\% & 16,792 & \textbf{4.80} & \textbf{29.45\%} \\
        DS4 & 2,414 & 1.61 & 24.64\% & 6,418 & 3.21 & 25.88\% & 16,784 & \textbf{4.80} & \textbf{29.44\%} \\
        DS5 & 1,655 & 1.10 & 16.89\% & 3,773 & 1.89 & 15.21\% & 11,677 & \textbf{3.34} & \textbf{20.48\%} \\
        DS6 & 1,727 & 1.15 & 17.63\% & 2,925 & 1.46 & 11.79\% & 11,262 & \textbf{3.22} & \textbf{19.75\%} \\
        DS7 & 2,982 & 1.99 & 30.44\% & 7,553 & 3.78 & \textbf{30.45\%} & \textbf{16,409} & \textbf{4.69} & 28.78\% \\
        DS8 & 1,865 & 1.24 & 19.04\% & 3,562 & 1.78 & 14.36\% & 11,773 & \textbf{3.36} & \textbf{20.65\%} \\
        \midrule
        Mean & 2,704 & 1.80 & 27.61\% & 6,471 & 3.24 & 26.09\% & \textbf{17,109} & \textbf{}\textbf{4.89} & \textbf{30.01\%} \\
        \bottomrule
    \end{tabular}}
    \label{tab:rq1_main}
\end{table}

After testing the eight dialogue systems, the results are presented in Table~\ref{tab:rq1_main}. MORTAR is capable of detecting 532\% more bugs than QAAsker and 164\% more bugs than METAL, with 15.0\% and 8.69\% higher turn failure rate than METAL and QAAsker respectively. In general, MORTAR shows higher effectiveness than the baselines.

\subsubsection{Efficiency} 
The metric bug per test case measures the bug density in testing. The higher this metric is, the more bugs the testing method can reveal through one test case execution, thus is more efficient. On average, MORTAR's test efficiency is 172\% higher than QAAsker and 51\% higher than METAL. 

\subsubsection{Robustness} 
When testing different dialogue systems, the turn failure rate (Rate\textsuperscript{+}) of QAAsker dropped by 33.37 percentage points (pp) from 51.42\% (DS1) to 17.63\% (DS6). METAL dropped by 40.38 pp from 52.17\% (DS1) to 11.79\% (DS6). MORTAR dropped by 32.89 pp from 52.64\% (DS1) to 19.75\% (DS6), the performance drop is lower than that of the best baseline. We then measure the coefficient of variation (CV) of performance when testing different dialogue systems. In Table~\ref{tab:rq1_cv}, the 10.3\% lower CV than best-performed baseline QAAsker indicates better robustness of MORTAR in testing. Collectively, it is observed that MORTAR has better robustness than the baseline in testing different dialogue systems. 

\begin{table}
    \centering
    \caption{The coefficient of variation (CV) of Rate\textsuperscript{+} when testing different dialogue systems. Lower is better.} 
    \adjustbox{max width=\linewidth}{%
    \begin{tabular}{l|rrr}
    \toprule
        & QAAsker & METAL & MORTAR \\
        \midrule
       CV & 0.389 & 0.475 & \textbf{0.349} \\
       \bottomrule
    \end{tabular}}
    \label{tab:rq1_cv}
\end{table}

\begin{tcolorbox}[colback=gray!10, colframe=black, boxrule=0.5pt]
\textbf{Answer to RQ1:} MORTAR is 164\% more effective, 51\% more efficient, and 10.3\% more robust than the single-turn MT baseline methods.
\end{tcolorbox}

\subsection{RQ2. Quality of Detected Bugs}
To answer RQ2, we examine the quality of bugs in terms of their correctness, diversity and uniqueness.

\subsubsection{Precision}
\label{sec:rq2.1}
To measure the precision of detected bug, we randomly sample 120 bugs detected by each testing method, such that a total of 360 bugs are manually checked. Two authors first independently give binary judgement of the correctness of each dialogue system output, then discuss uncertain cases and, where possible, reach a consensus on the final decision. A bug is classified as false positive if both evaluators agree that \textit{a)} the input turn question is clear and understandable, and \textit{b)} the dialogue system’s output is factually correct, aligns with the answer requirements, and has only reasonably limited additional content if there is any. Otherwise, the bug will be classified as a true positive.

\begin{table}[]
    \centering
    \caption{The manual check results of precision of sampled bugs detected by different methods.}
    \begin{tabular}{l|rrr}
    \toprule
    & QAAsker & METAL & MORTAR \\
    \midrule
    Precision & 49.2\% & 60.8\% & \textbf{75.8\%} \\
    Cohen's Kappa  & 0.716 & 0.683 & 0.733 \\
    \bottomrule
    \end{tabular}
    \label{tab:rq21_precision}
\end{table}

As described in Table~\ref{tab:rq21_precision}, MORTAR realises 24.6\% higher precision in bug detection than the best-performing baseline METAL, with a higher Cohen's Kappa than baselines. This indicates a higher level of inter-rater agreement on MORTAR’s detection outcomes. The lower Kappa score for METAL may reflect a greater degree of unresolved disagreement between the two evaluators. Further discussion of false positive detections is provided in Section \ref{result:disciussion:FPR}. Overall, MORTAR demonstrates a higher proportion of true positive detections, indicating better precision compared to the baselines.

\subsubsection{Diversity}

\begin{table}
    \centering
    \caption{The number of different types of bugs and the coefficient of variation (CV) of average number of different types of bug. The lowest CV indicates bugs revealed by MORTAR are of higher diversity.}
     \adjustbox{max width=\linewidth}{%
    \begin{tabular}{l|rrr|rrr|rrr}
        \toprule
        \multirow{2}{*}{DS} & \multicolumn{3}{c}{QAAsker} & \multicolumn{3}{c}{METAL} & \multicolumn{3}{c}{MORTAR} \\
        \cmidrule(lr){2-4} \cmidrule(lr){5-7} \cmidrule(lr){8-10}
        & L1-Bugs & L2-Bugs & L3-Bugs & L1-Bugs & L2-Bugs & L3-Bugs & L1-Bugs & L2-Bugs & L3-Bugs \\
        \midrule
        DS1 & 2,248 & 2,758 & 31 &7,717 & 5,149 & 74 & 15,468 & 14,254 & 287 \\
        DS2 & 927 & 2,248 & 224 &3,870 & 4,215 & 420 & 7,731 & 12,908 & 1,527 \\
        DS3 & 420 & 1,690 & 448 &2,425 & 2,898 & 770 & 4,687 & 9,125 & 2,980 \\
        DS4 & 518 & 1,748 & 148 &2,624 & 3,511 & 283 & 5,762 & 10,286 & 736 \\
        DS5 & 119 & 883 & 653 &1,018 & 1,720 & 1,035 & 1,950 & 5,939 & 3,788 \\
        DS6 & 87 & 928 & 712 &981 & 1,197 & 747 & 1,909 & 5,560 & 3,793 \\
        DS7 & 422 & 2,129 & 431 &2,405 & 4,247 & 901 & 4,781 & 9,816 & 1,812 \\
        DS8 & 126 & 938 & 801 &900 & 1,508 & 1,154 & 1,870 & 5,640 & 4,263 \\
        \midrule
        Mean & 608 & 1,665 & 431 &2,742 & 3,055 & 673 & 5,519 & 9,191 & 2,398 \\
        \midrule
        CV & & 0.604 & & & 0.490 & & & \textbf{0.487} & \\
        \bottomrule
    \end{tabular}}
    \label{tab:bug_type_compare}
\end{table}
Diversity indicates the testing method's capability of diversified usage of the original dataset to detect bugs. The definition of L1-Bugs, L2-Bugs and L3-Bugs are given in Section~\ref{sec:criteria}. According to the results in Table~\ref{tab:bug_type_compare}, on average, MORTAR reveals a significantly higher number of different types of bugs while keeping a balance of different types of bugs. The lowest coefficient of variation (CV) of the average number of bugs of different types indicates better diversity of the bugs detected by MORTAR.

\subsubsection{Uniqueness}

\begin{figure}[!t]
    \centering
    \includesvg[width=1\linewidth]{figs/RESULTS/RQ2.3.fig.svg}
    \caption{Seed question overlap of detected bugs.}
    \label{fig:rq2_overlap}
\end{figure}

To measure the uniqueness of bugs, we collect the seed question of bugs that are not revealed in RBT but by MORTAR and baseline methods. The seed question overlap is presented in Fig~\ref{fig:rq2_overlap}. On average, 32.9\% of the bugs revealed by QAAsker are unique, 26.9\% of the bugs revealed by METAL are unique, while over 42.3\% of the bugs revealed by MORTAR are unique, which is higher than all baselines.  

\begin{tcolorbox}[colback=gray!10, colframe=black, boxrule=0.5pt]
\textbf{Answer to RQ2:} MORTAR is 24.6\% more precise than baseline methods. The bugs revealed by MORTAR are more diverse, and over 42.3\% of the revealed bugs are unique.
\end{tcolorbox}

\subsection{RQ3. Ablation Study}
In MORTAR, seven dialogue-level perturbations are adopted to generate metamorphic test cases, and four MRs are used to form multi-turn metamorphic testing for dialogue systems. This RQ focuses on the contribution of each component to the overall effectiveness in test case generation and bug detection.

\subsubsection{Ablation study on MR effectiveness}
\begin{table}[]
    \centering
    \caption{The number of bugs detected by MORTAR when a certain MR is excluded. For example, the column MR1 indicates the number of detected bugs without using MR1 but only using MR2, MR3 and MR4, and the percentage in the bracket indicates the decrease ratio when compared with original performance using the full set of MRs. The bold values indicate the most effective MR when testing different dialogue systems.}
    \adjustbox{max width=\linewidth}{%
    \begin{tabular}{l|rrrrr}
    \toprule
        & Original & MR1 & MR2 & MR3 & MR4 \\
        \midrule
    DS1 & 30,009 &  8,459 (\textbf{-71.81\%}) & 27,524 (-8.28\%) & 25,850 (-13.86\%) & 28,194 (-6.05\%) \\
    DS2 & 22,166 &  8,888 (\textbf{-59.90\%}) & 18,657 (-15.83\%) & 18,621 (-15.99\%) & 20,332 (-8.27\%) \\
    DS3 & 16,792 &  8,093 (\textbf{-51.80\%}) & 12,620 (-24.85\%) & 14,794 (-11.90\%) & 14,869 (-11.45\%) \\
    DS4 & 16,784 &  7,939 (\textbf{-52.70\%}) & 12,928 (-22.97\%) & 14,685 (-12.51\%) & 14,800 (-11.82\%) \\
    DS5 & 11,677 &  7,492 (-35.84\%) & 6,919 (\textbf{-40.75\%}) & 10,853 (-7.06\%) & 9,767 (-16.36\%) \\
    DS6 & 11,262 &  7,416 (-34.15\%) & 6,418 (\textbf{-43.01\%}) & 10,618 (-5.72\%) & 9,334 (-17.12\%) \\
    DS7 & 16,409 &  7,874 (\textbf{-52.01\%}) & 12,343 (-24.78\%) & 14,495 (-11.66\%) & 14,515 (-11.54\%) \\
    DS8 & 11,773 &  7,666 (-34.88\%) & 6,737 (\textbf{-42.78\%}) & 11,083 (-5.86\%) & 9,833 (-16.48\%) \\
    \midrule
    Mean & 17,109 &  7,978 (\textbf{-53.37\%}) & 13,018 (-23.91\%) & 15,125 (-11.60\%) & 15,206 (-11.13\%) \\
    \bottomrule
    \end{tabular}}
    \label{tab:ablation_mr}
\end{table}

To measure the effectiveness of the MRs, we perform an ablation study by collecting the number of detected bugs when each of the MRs is disabled one at a time. According to the result shown in Table~\ref{tab:ablation_mr}, among all MRs, MR1 generally contributes most to the overall effectiveness. MR2 shows unique major contribution when testing certain dialogue systems (e.g., DS5, DS6 and DS8). In general, all MRs contribute to the overall effectiveness, and the removal of any MR will cause an overall performance drop by over 11\%.

\subsubsection{Ablation study on perturbation effectiveness}

\begin{table*}[]
    \centering
    \caption{The number of bugs detected by MORTAR without each perturbation. For example, the column RS indicates the number of detected bugs without using perturbation RS, and the percentage in the bracket indicates the decrease ratio. The \textbf{bold} values indicate the most effective perturbation, and the \textit{italic} values indicate the least effective perturbations.}
    \adjustbox{max width=\linewidth}{%
    \begin{tabular}{l|rrrrrrrr}
    \toprule
    & Original & RS & RR & RD & RSR & RSD & RRD & RSRD \\
    \midrule
    DS1 & 30,009 &  25,392 (-15.39\%) & 27,978 (\textit{-6.77\%}) & 24,618 (-17.96\%) & 27,297 (-9.04\%) & 24,127 (\textbf{-19.60\%}) & 27,047 (-9.87\%) & 26,517 (-11.64\%) \\
    DS2 & 22,166 &  18,740 (-15.46\%) & 20,660 (\textit{-6.79\%}) & 18,180 (-17.98\%) & 20,196 (-8.89\%) & 18,003 (\textbf{-18.78\%}) & 19,912 (-10.17\%) & 19,628 (-11.45\%) \\
    DS3 & 16,792 &  14,207 (-15.39\%) & 15,640 (\textit{-6.86\%}) & 13,794 (-17.85\%) & 15,289 (-8.95\%) & 13,504 (\textbf{-19.58\%}) & 15,112 (-10.00\%) & 14,789 (-11.93\%) \\
    DS4 & 16,784 &  14,018 (-16.48\%) & 15,653 (\textit{-6.74\%}) & 13,978 (-16.72\%) & 15,142 (-9.78\%) & 13,596 (\textbf{-18.99\%}) & 15,223 (-9.30\%) & 14,755 (-12.09\%) \\
    DS5 & 11,677 &  9,757 (-16.44\%) & 10,901 (\textit{-6.65\%}) & 9,754 (-16.47\%) & 10,515 (-9.95\%) & 9,344 (\textbf{-19.98\%}) & 10,552 (-9.63\%) & 10,275 (-12.01\%) \\
    DS6 & 11,262 &  9,388 (-16.64\%) & 10,483 (\textit{-6.92\%}) & 9,436 (-16.21\%) & 10,147 (-9.90\%) & 8,995 (\textbf{-20.13\%}) & 10,223 (-9.23\%) & 9,877 (-12.30\%) \\
    DS7 & 16,409 &  13,802 (-15.89\%) & 15,311 (\textit{-6.69\%}) & 13,594 (-17.16\%) & 14,812 (-9.73\%) & 13,129 (\textbf{-19.99\%}) & 14,867 (-9.40\%) & 14,500 (-11.63\%) \\
    DS8 & 11,773 &  9,820 (-16.59\%) & 10,968 (\textit{-6.84\%}) & 9,862 (-16.23\%) & 10,634 (-9.67\%) & 9,413 (\textbf{-20.05\%}) & 10,647 (-9.56\%) & 10,325 (-12.30\%) \\
    \midrule
    Mean & 17,109 &  14,390 (-15.89\%) & 15,949 (\textit{-6.78\%}) & 14,152 (-17.28\%) & 15,504 (-9.38\%) & 13,764 (\textbf{-19.55\%}) & 15,448 (-9.71\%) & 15,083 (-11.84\%) \\
    \bottomrule
    \end{tabular}}
    \label{tab:ablation_pert}
\end{table*}

To measure the effectiveness of perturbations, we perform an ablation study by collecting the number of detected bugs when each of the perturbations is disabled one at a time. According to the Table~\ref{tab:ablation_pert}, perturbation RSD (dialogue round duplication and shuffle) contributes the most to overall effectiveness, with 19.55\% of bugs revealing relies on it. The remaining perturbations also contribute significantly to the testing process. On average, the removal of any perturbation will cause at least 6.78\% (column RR: dialogue round reduction) performance deterioration of MORTAR.

\subsubsection{Ablation study on the equivalent context check}

\begin{figure}
    \centering
    \includesvg[width=0.9\linewidth]{figs/RESULTS/RQ3.3.fig}
    \caption{The F1 score of equivalent context check results. Both decontextualised question similarity check and reference answer similarity check contribute to the overall effectiveness of equivalent context check.}
    \label{fig:rq33_ec}
\end{figure}

To measure the contribution of each component in equivalent context check, we randomly sample one dialogue from each perturbation, and a total of 141 rounds of questions are manually labelled with context equivalence results. The manual labels are compared against the performance of equivalent context without either decontextualised question similarity comparison or reference answer similarity comparison, and the results are presented in Fig~\ref{fig:rq33_ec}. Merely relying on decontextualised question similarity check or reference answer check cannot achieve the best performance. With both components, the overall F1 score reached 0.922, indicating high correctness of equivalent context check results. 

Additionally, the equivalent context check can also be implemented with a pure LLM-as-a-judge approach without the explicitly predefined processing workflow. We compare the effectiveness of the equivalent context check in MORTAR and the LLM-as-a-judge approach with the identical LLM (gpt-oss-120B) on the manually labelled dataset. The prompt template used in the pure LLM judge-based equivalent context check is provided in the supplementary material. According to the prompt, the LLM is instructed to think step by step and judge whether the target question with different contexts shall be answered with the same answer. The LLM-as-a-Judge approach and $EC(\cdot)$ in MORTAR are given the same input, and the inference temperatures are both 0.5. According to the results shown in Table~\ref{tab:llmasajudgecomp}, $EC(\cdot)$ in MORTAR achieves a higher F1 score than the LLM-as-a-Judge approach. Within the sampled dataset, the LLM-as-a-Judge approach tends to give more judgement that the two given contexts are ``equivalent'' and less judgement of ``inequivalent'', resulting in high recall on the ``equivalent'' category while very low recall on the ``inequivalent'' category. As a comparison, $EC(\cdot)$ in MORTAR can identify such situations and give correct judgment of ``inequivalent'', indicating less bias and higher accuracy on the equivalent context check process. In general, $EC(\cdot)$ in MORTAR outperforms the pure LLM-as-a-judge baseline, the roles of LLM in $EC(\cdot)$ are constrained to executing specific natural language processing tasks within the explicitly predefined processing workflow, rather than making the final verdict directly, such that mitigates the bias of the LLM towards the ``equivalent'' category, which ensures the effectiveness of the equivalent context check.

\begin{table}[]
  \centering
  \caption{Performance Comparison between equivalent context check ($EC(\cdot)$) in MORTAR and the approach based on LLM-as-a-Judge. The precision and recall are calculated on both manual labels of equivalent and inequivalent  respectively, the Micro-F1 score (F1 Score) indicates the overall correctness in equivalent context check. The bold values indicates better performance.}
  \label{tab:llmasajudgecomp}
  \adjustbox{max width=\linewidth}{
  \begin{tabular}{l|rr|rr|r}
    \toprule
    & \multicolumn{2}{c}{Equivalent} & \multicolumn{2}{c}{Inequivalent} & \multirow{2}{*}{F1 Score} \\
    \cmidrule(lr){2-3} \cmidrule(lr){4-5}
    & Precision & Recall & Precision & Recall & \\
    \midrule
    $EC(\cdot)$ in MORTAR & \textbf{0.949} & 0.957 & \textbf{0.792} & \textbf{0.760} & \textbf{0.922} \\
    LLM-as-a-Judge & 0.827 & \textbf{0.991} & 0.500 & 0.040 & 0.823 \\
    \bottomrule
  \end{tabular}}
\end{table}

To verify the robustness of equivalent context check under normal LLM variability, we repeat equivalent context check for 20 times on the manually labelled dataset and report how often the results flips between ``equivalent'' and ``inequivalent''. Following the practice in related research\cite{wang2023selfconsistency}, the LLM inference temperature is setup to be 0.5. According to the experimental results, over 84\% judgements given by equivalent context check are identical with the first attempt, and Fleiss' Kappa \cite{vashurin2025benchmarking,correia2026comparison} reach 0.63 ($\geq0.6$: substantial agreement). These results indicate that while normal LLM variability does introduce minor fluctuations, the final binary decisions (``equivalent'' or ``inequivalent'') made by equivalent context check demonstrate substantial agreement across multiple repeats. Therefore, the equivalent context check in MORTAR is regarded as robust under normal LLM variability.

It is worth noting that the bug counts reported in this section refer to MR violations rather than manually verified faults. As reported in Section~\ref{sec:rq2.1} (Table~\ref{tab:rq21_precision}), 75.8\% of MORTAR's detections are confirmed as genuine faults by manual inspection. The ablation results above can be scaled by this factor to obtain conservative estimates of true bug counts.

\begin{tcolorbox}[colback=gray!10, colframe=black, boxrule=0.5pt]
\textbf{Answer to RQ3:} All perturbations and MRs contribute to the effectiveness of test case generation and bug discovery. Every perturbation and MR accounts for more than 6.78\% and 11.13\% to the overall bug detection performance respectively. Each component in equivalent context check contributes to the overall performance, and the equivalent context check is robust under normal LLM variability.
\end{tcolorbox}

\subsection{RQ4: Sensitivity analysis}
\label{sec:RQ4}
In this RQ, we analyse the sensitivity of MORTAR to different perturbation ratio and semantic similarity thresholds.

\subsubsection{Sensitivity to perturbation ratio}

\begin{figure}[!ht]
    \centering
    \includesvg[width=\linewidth]{figs/RESULTS/RQ4.1.fig}
    \caption{Sensitivity to perturbation ratio.}
    \label{tab:rq4_pr}
\end{figure}

To analyse the sensitivity to the perturbation ratio, we chose DS5 as the subject as it shows medium performance. Perturbation ratios ranging from 0.1 to 0.5 with a step of 0.2 were included for comparison. According to the results shown in Fig.~\ref{tab:rq4_pr}, no significant performance variation of MORTAR was observed. As the perturbation ratio rises, the number of perturbed questions and bugs show a minor decrease, but the turn failure rate (Rate\textsuperscript{+}) is insensitive to different perturbation ratios. Moreover, the distribution of different types of bugs and unique bugs shows no major changes. It can be concluded that MORTAR's performance is robust in the tested perturbation ratio range.

\subsubsection{Sensitivity to equivalent context check threshold}
The sensitivity analysis result of different equivalent context check threshold is shown in Fig.~\ref{fig:rq33_ec}. It is observed that with overall threshold 0.8, equivalent context check reaches its best measured performance. This corresponds with the design of equivalent context check. Some questions might contain redundant information and caused the similarity to be lower than other pairs. Reference answer check functions as an auxiliary check to allow those cases to be correctly judged. MORTAR is sensitive to equivalent context check threshold, the optimal performance is observed at threshold 0.8.

\begin{tcolorbox}[colback=gray!10, colframe=black, boxrule=0.5pt]
\textbf{Answer to RQ4:} MORTAR is insensitive to perturbation ratio in selected value range, but sensitive to the equivalent context check threshold where 0.8 is determined as the optimal value..
\end{tcolorbox}

\subsection{Discussion}

\subsubsection{Case study: the false positive detections}
\label{result:disciussion:FPR}
The most common false positive detections are overlong answers. For example, the short answer ``Yes'' is expected in yes-no questions, the prompt also requests the answer to be concise, exact and short. As some questions might require detailed analysis, the DS tend to explain the answer with overlong paragraphs, as a result, the semantic similarity is lower than the threshold. The labellers report that some reasonably wordy outputs of dialogue systems could be regarded as false positive detections, e.g. the expected answer: ``In their home'' and the output: ``Mike McLelland and his wife Cynthia were found in their home in Kaufman County, Texas.'' do not have high semantic similarity (0.56), but the output is actually acceptable. Nevertheless, considerable outputs exceed the necessary explanation and give exceedingly long paragraphs, they are regarded as true positives. We expect future improvement in semantic similarity measurement to reduce false positive rates.

Besides, during manual check, both participants reported that METAL's perturbation brings too much noise to the original question, and considerable questions' semantics have been altered, which contradicts the expectation. For example, in multi-turn testing with the theme of biology, a question \textit{``What is a gene?} is perturbed with \textit{add random word} perturbation in METAL, and the perturbed question \textit{``What is Pear a gene?''} significantly changed the semantics of the sentence. The dialogue system uses its internal knowledge to answers this question with explanations about the genes in fruits. Besides, the synonym replacement cannot guarantee robust behaviours. For example, the question ``Who is in charge?'' is perturbed into ``World-Health-Organisation is in charge?'' where ``Who'' is replaced by ``World-Health-Organisation'' which is improper.

The common issue with QAAsker is the quality of generated test cases. The rule-based sentence analysis in QAAsker cannot cover situations beyond the predefined rules, especially when ellipsis phenomenon is frequently observed in multi-turn dialogue. For example, the question ``What was he doing before his death?'' with the original answer ``returning from a convenience store.'' is reformed with QAAsker MR1 into another wh-question ``When was he returning from a convenience store?'' and expects the answer ``his death''. DS8 answers the perturbed question with ``February 26, 2012'' which is verifiable in reference text and shows strong reasoning capability, this bug is regarded as a typical false positive.

\subsubsection{Case study: the root cause of bugs}
The categorisation of failures is based on the participants' reported observations in manual check in Section~\ref{sec:rq2.1}. The common types of failures are: a) hallucinated content, b) failed prompt instruction following, c) imperfect context reasoning and d) mistakenly triggered guardrail. The hallucinated content is the situation that the DS outputs content with information never mentioned in the reference document and context. For example, the question ``Where did Cotton's mother put Cotton to clean the paint off?'' should be answered with ``a bucket of water'', while DS1 outputs ``It's possible for Cotton's mommy possibly rubbed her face on cotton to clean it off.'' which cannot be verified by the reference story. For failed prompt instruction following, all DS are required to provide ``concise and short'' answer without explanation, while from DS1 to DS8 all occasionally provide unnecessary analysis or ``Explanation'' tagged content. The imperfect context reasoning is the phenomenon that the DS cannot correct the misunderstanding in context and follows a wrong dialogue trajectory. It is more common for DS1 and DS2 whose performance are relatively weaker in testing. The mistakenly triggered guardrail is only observed with DS7. In the whole testing, 27 questions are rejected by DS7 as they are flagged as potentially violating the usage policy, while same questions are successfully answered by DS8. We anticipate the root cause to be ill-suited guardrail of GPT-5-nano model.

\subsubsection{On the performance evaluation of dialogue systems}
As depicted in Section~\ref{sec:test_objects}, our experiments are conducted on eight dialogue systems that are based on different LLMs in terms of source and parameter size. DS1, DS2 and DS3 are based on LLMs of the same source but different parameter sizes, they form the comparison group 1. DS3, DS4, DS5 and DS6 are based on different sourced models with similar parameter sizes (7-9 billion), they form the comparison group 2. DS7 and DS8 are based on closed-source LLMs and form the comparison group 3. In group 1 and group 3, a scaling trend is observed through the trend of Rate\textsuperscript{+}. Larger models exhibit fewer bugs and a lower Rate\textsuperscript{+}. In group 2, the performance of different DS is at a similar level. To compare between open-sourced LLMs and closed-source LLMs, the GPT-5 LLM that DS8 is based on, is anticipated with larger parameter size than the tested open-source LLMs. GPT-5 model performs well on many benchmarks\cite{hendrycks2020mmlu, jimenez2023swebench}, while it is not performing significantly better than other DS in multi-turn MT. Larger language models are typically capable of strong memory of knowledge in the training dataset and better reasoning capability \cite{brown2020language}. However, according to the test results in Table \ref{tab:ablation_mr}, LLM with a larger parameter size exhibits higher proportions of violations to MR2 and MR4. This contradicts the trend of increased reasoning capability as expectations. While the exact architectural limitations of GPT-5 model remain unclear, the results indicate that highly scaled models may still struggle with complex contextual perturbations. MORTAR is capable of reusing this dataset and building multi-turn metamorphic testing to reveal new bugs. Additionally, we expect the integration of MORTAR in model training to be beneficial to avoid over-fitting and improve the generalisation in real-world applications.

\subsubsection{On the mitigation method}

\begin{table}[]
    \centering
    \caption{The performance of high reasoning effort (DS9) and low reasoning effort (DS7) when tested with MORTAR.}
    \begin{tabular}{l|rrrr}
    \toprule
        & Bugs & B/TC & Rate\textsuperscript{+} & Output tokens \\
    \midrule
    DS7 & 16,409 & 4.69 & 28.78\% & 4.59M  \\
    DS9 & 14,316 & 4.56 & 25.11\% & 37.99M \\ 
    \bottomrule
    \end{tabular}
    \label{tab:dis4_re}
\end{table}

A potential method that enhances the capability of LLM-based dialogue systems is to increase the reasoning effort of backbone LLMs \cite{wei2022chain}. We configure the reasoning effort of the GPT-5-nano model to high and form DS9, all other settings are kept the same as DS7. DS9 is tested with MORTAR using default parameters in RQ1. According to the results shown in Table~\ref{tab:dis4_re}, when configured with a higher reasoning effort of the backbone LLM, the number of bugs is decreased by 12.76\%. Meanwhile, the number of output tokens is increased by over seven times. The additional output tokens form the reasoning content, resulting in fewer bugs in testing. Increasing the reasoning effort of the backbone LLM of the dialogue system appears to be an effective mitigation method. Besides, the detected bugs are potentially helpful to improve the performance of LLM-based dialogue systems. The bugs can also be regarded as negative samples, and using the outputs that violates MRs (e.g., MR1 and MR2) as negative preferences, post-training approaches\cite{mao2025as,zhang2024negative} can finetune the LLM to learn from the mistakes and realise better performance. We expect this future work to explore this direction and improve LLM-based dialogue systems with bugs revealed by MORTAR.

\section{Threats to Validity} 
\label{sec:validity}
\subsubsection{Perturbation and equivalent context check}
The proposed dialogue-level perturbations rely on randomness to generate follow-up test cases, which cannot gain fine-grained control over the actual effect on follow-up test cases without scenario-specific designs. Therefore, an equivalent context process check is proposed to judge the actual effect of perturbation and guide the selection of proper MR. The framework of MORTAR enables fully automated metamorphic testing, while it might introduce wrong judgments and cause false positive detections. As there does not exist any plug-and-play tool that is primarily designed for the purpose of equivalent context checking, we crafted a feasible checking process to automate MORTAR. We further conduct a sensitivity analysis of the threshold and an ablation study in the equivalent context check to verify the framework design, and manually check the bugs to minimise the threats to validity.

\subsubsection{Generalisability of results}
This study assesses MORTAR using the multi-domain multi-turn QA dataset and eight LLMs against single-turn MT baselines. The dataset selection is aligned with the scope of this study, while taxonomies of dialogue in both linguistics and natural language processing are much broader in scope and sophisticated in category. When using different task-specific and domain-specific seed datasets, e.g., code generation, or testing LLMs not included in this study, the effectiveness of MORTAR may vary. Although we aim to include diverse models and scenarios, the results might not fully generalise to other domains. We endeavour to extend MORTAR to a broader scope in future work.

\section{Related Work}
\label{sec:LR}
\subsection{Testing LLM-based Dialogue Systems}
Current dialogue test cases are generally from human-generated datasets and LLM-generated datasets. SQuAD and SQuAD 2.0 are classic single-turn QA datasets that have been widely used for training and testing of QA systems \cite{rajpurkar2016squad1, rajpurkar2018squad2}. CoQA dataset is a free-form multi-turn conversational QA dataset with unanswerable questions \cite{reddy2019coqa}. MuTual is a multi-turn reasoning-based dialogue dataset in open-domain \cite{cui2020mutual}. MT-Bench is a 2-turn open-end dialogue dataset that is used to evaluate chatbots' conversation performance \cite{zheng2023llmasajudge}. The ``LLM-as-a-judge'' approach is reported as feasible when comparing LLMs' alignment with human preference, and using LLM to generate test cases became a recent trend \cite{sun2024parrot, bai2024mt101, kwan2024mteval, wang2023mint, duan2024botchat}. However, it is worth noting that the ``LLM-as-a-judge'' approach comes with intrinsic shortcomings \cite{zheng2023llmasajudge, huang2024empirical}, and the evaluations are restricted to LLM judge scores but not pass-fail criteria. Besides, training data contamination is receiving increasing attention \cite{mirzadeh2024gsm}. The area calls for more complete and reliable testing for LLM-based dialogue systems.

\subsection{Metamorphic Testing}
To alleviate the oracle problem in software testing, MT was first introduced in 1998 \cite{chan1998application}. Later and up to now, MT is commonly used in both traditional software systems \cite{zhuang2023testing, zhou2015metamorphic} and machine learning systems \cite{xie2011MT, chen2021qaasker, wang2023mttm}. The advantages of MT are simplicity in concept, the straightforward implementation, ease of automation, and low cost\cite{chen2018metamorphic}. These lead to the potential of effective testing for complicated systems, e.g., the compiler\cite{le2014compiler} the black-boxed GenAI systems\cite{genaist}. Specifically, METAL uses utterance perturbations and MR templates to evaluate LLMs \cite{hyun2024metal}. Ontology-based MT provides concrete task-oriented test cases but falls short in generalisation\cite{bovzic2022ontology}. KGIT uses knowledge graphs as test seeds and builds a metamorphic testing-based scheme to implement large-scale inference tests \cite{wang2024kgit}. Drowzee uses metamorphic testing and constrained logic programming to detect hallucinations in LLMs \cite{li2024halluvault}. DialTest uses a set of MRs to effectively generate test cases for dialogue systems \cite{liu2021dialtest}. Existing MT approaches lack systematic analysis and usage of dialogue-level perturbations and multi-turn dialogue test case generation. Multi-turn dialogue testing remains challenging.

\section{Conclusion and future works}
\label{sec:conclusion}
LLM-based dialogue systems are now widely used in real-world applications. While there exist many single-turn testing methods, they are limited to the single‑turn scenarios, leaving the real‑world multi‑turn usage scenario significantly underexplored. In this research, a multi-turn metamorphic testing approach, MORTAR, is proposed to mitigate the oracle problem in dialogue system testing. In MORTAR, a series of formalised MRs and dialogue-level perturbations are proposed and implemented with the multi-turn MT framework to generate follow-up test cases and reveal bugs in LLM-based dialogue systems. Given an automated MR matching mechanism, MORTAR can be fully automated. The test results show MORTAR is more effective than the most effective single-turn metamorphic testing baseline. MORTAR detects over 164\% more bugs than the single-turn MT baseline, and over 42\% of them are unique bugs. The revealed bugs are of higher quality when compared with the baseline in terms of diversity, precision and uniqueness. The component contribution analysis further validates the effectiveness of components in MORTAR.

We wish to point out that \textit{(i)} these four types of MRs are not restricted to the seven types of perturbations introduced in this paper, and \textit{(ii)} different application domains for the multi-turn dialogue systems may have different MRs as their characterisations. MORTAR is expected to inspire both dialogue system developers and researchers to build more comprehensive testing methods for dialogue systems. In the future, we intend to refine the equivalent context check process and reduce the false positive detections. The semantic similarity measurement can also be enhanced to improve the accuracy of judgment. Besides, the detected bugs can be further used as negative training sets, and pave the way for improving LLM-based dialogue systems in future work.

\bibliographystyle{IEEEtran}
\bibliography{ref}

\end{document}


\title{}

\maketitle
\section*{Supplementary material for MORTAR}

\subsection{Prompts used in equivalent context check}
\begin{promptbox}[Reference material topic extraction]
\footnotesize\ttfamily
System: \\
You are an expert in linguistics. You will help the user to finish their task precisely. Do not give any explanation and apologize.\\
\tcbline
Input: \\
Task: given an article, extract a concise topic that best describes the article in a sentence. Only output the topic without any additional text.
\\=================\\
Article: \\
\texttt{\{ ARTICLE\_TEXT \}}
\\=================\\
Output:
\end{promptbox}

\begin{promptbox}[Decontextualisation]
\footnotesize\ttfamily
System: \\
Task: Rewrite the next question so it can be understood by itself without looking at the rest of the dialogue.\\=================\\Instructions:\\- In the given conversation between a student who asks questions about an article about the topic of \{ ARTICLE\_TOPIC \} and a teacher who is answering the questions, most questions can be understood as part of the ongoing dialogue. Edit the understandable question so that it makes sense by itself without looking at the rest of the dialogue. Think how you would ask the same question to someone who has no idea about what the dialogue is about and has never seen any of the previous utterances and still get the same answer. \\\\- Questions should stay in a natural-sounding English after editing. Do not use new words other than those in the conversation unless it is necessary. Copy from previous questions, answers and conversation topics as often as possible. When editing a question, keep its original structure whenever possible. Do not come up with a different or better question. Stick to the given wording. It is okay to have pronouns in an edit as long as the edit has enough information about what the pronouns refer to. For example 'Where was Carroll when she joined the American Party?' is a perfectly fine edit.\\\\- Pay special attention to questions that ask if there is anything/anyone else or asking for telling more about something. It is important to mention in your edit “anything/anyone else besides what/ whom”.\\\\- If a question is associated with a question asked in context, \\\\- If a question can be understood without looking at the rest of the conversation, output the ORIGINAL question. Do NOT make unnecessary edits. Try to keep the edits short as long as they make sense each by itself. Although typos are quite rare, please fix any typos you encounter.\\\\- If a question CANNOT be understood under current context questions, output \#unclear\#.\\\\- Just output the edited question for independently understandable question, or  \#unclear\# for unclear question\\
\tcbline
Input: \\
Article:\\
\{ ARTICLE\_TEXT \}\\
=================\\
Context questions:\\
\{ CONTEXT\_QUESTIONS \}\\
=================\\
Next question to decontextualise:\\
\{ TARGET\_QUESTION \}\\
=================\\
Output:
\end{promptbox}

\begin{promptbox}[Testing LLM-based dialogue systems]
\footnotesize\ttfamily
System: \\
Read the following story:\\
\{ STORY\_CONTENT \}\\
I'm going to ask you some questions about the story. Please give me concise and short answer WITHOUT explanation. Output unknown if you don't know the answer.\\
\tcbline
Input: \\
\{ DIALOGUE\_HISTORY \}\\
\{ TARGET\_QUESTION \}
\end{promptbox}

\begin{promptbox}[Example of test input]
\footnotesize\ttfamily
System/User: \\
Read the following story:\\\\
Once upon a time, in a barn near a farm house, there lived a little white kitten named Cotton. Cotton lived high up in a nice warm place above the barn where all of the farmer's horses slept. But Cotton wasn't alone in her little home above the barn, oh no. She shared her hay bed with her mommy and 5 other sisters. All of her sisters were cute and fluffy, like Cotton. But she was the only white one in the bunch. The rest of her sisters were all orange with beautiful white tiger stripes like Cotton's mommy. Being different made Cotton quite sad. She often wished she looked like the rest of her family. So one day, when Cotton found a can of the old farmer's orange paint, she used it to paint herself like them. When her mommy and sisters found her they started laughing. \\
"What are you doing, Cotton?!" \\
"I only wanted to be more like you". \\
Cotton's mommy rubbed her face on Cotton's and said "Oh Cotton, but your fur is so pretty and special, like you. We would never want you to be any other way". And with that, Cotton's mommy picked her up and dropped her into a big bucket of water. When Cotton came out she was herself again. Her sisters licked her face until Cotton's fur was all all dry. \\
"Don't ever do that again, Cotton!" they all cried. "Next time you might mess up that pretty white fur of yours and we wouldn't want that!" \\
Then Cotton thought, "I change my mind. I like being special".\\\\
I'm going to ask you some questions about the story. Please give me concise and short answer WITHOUT explanation. Output unknown if you don't know the answer.\\
\tcbline
Turn 1: \\
User: Whose paint was it?\\
Assistant: the farmer\\
\tcbline
Turn 2:\\
User: Was Cotton happy that she looked different than the rest of her family?\\
Assistant: no\\
\tcbline
Turn 3:\\
User: Did she live alone?\\
\end{promptbox}

\subsection{Prompts used in equivalent context check based on straightforward LLM-as-a-judge approach}
\begin{promptbox}[Equivalent context check (pure LLM-as-a-judge)]
\footnotesize\ttfamily
You are a helpful assistant.\\
You are given a story and two lists of questions. Your task is to judge whether the last question of the two lists should be answered with the same answer. If the contexts of the two questions are equivalent, then they should be answered with the same answer, and you should answer with "yes", otherwise answer with "no".
\\=================\\
Article: \{story\} \\
Question sequence 1: \{QUESTION\_SEQ\_1\}\\
Question sequence 2: \{QUESTION\_SEQ\_2\}
\\=================\\
Please think step by step and give your final answer and output in json format\{``answer'': ``yes'' or ``no''\}\\
\end{promptbox}

\subsection{Example of test cases}

\begin{promptbox}[Reference material and original QA]
\footnotesize\ttfamily
Once upon a time, in a barn near a farm house, there lived a little white kitten named Cotton. Cotton lived high up in a nice warm place above the barn where all of the farmer's horses slept. But Cotton wasn't alone in her little home above the barn, oh no. She shared her hay bed with her mommy and 5 other sisters. All of her sisters were cute and fluffy, like Cotton. But she was the only white one in the bunch. The rest of her sisters were all orange with beautiful white tiger stripes like Cotton's mommy. Being different made Cotton quite sad. She often wished she looked like the rest of her family. So one day, when Cotton found a can of the old farmer's orange paint, she used it to paint herself like them. When her mommy and sisters found her they started laughing. \\

"What are you doing, Cotton?!" \\

"I only wanted to be more like you". \\

Cotton's mommy rubbed her face on Cotton's and said "Oh Cotton, but your fur is so pretty and special, like you. We would never want you to be any other way". And with that, Cotton's mommy picked her up and dropped her into a big bucket of water. When Cotton came out she was herself again. Her sisters licked her face until Cotton's fur was all all dry. \\

"Don't ever do that again, Cotton!" they all cried. "Next time you might mess up that pretty white fur of yours and we wouldn't want that!" \\

Then Cotton thought, "I change my mind. I like being special".\\

\tcbline
Q1: What color was Cotton?\\
A1: white
\tcbline
Q2: Where did she live?\\
A2: in a barn
\tcbline
Q3: Did she live alone?\\
A3: no
\tcbline
Q4: Who did she live with?\\
A4: with her mommy and 5 sisters
\tcbline
Q5: What color were her sisters?\\
A5: orange and white
\tcbline
Q6: Was Cotton happy that she looked different than the rest of her family?\\
A6: no
\tcbline
Q7: What did she do to try to make herself the same color as her sisters?\\
A7: she painted herself
\tcbline
Q8: Whose paint was it?\\
A8: the farmer
\tcbline
Q9: What did Cotton's mother and siblings do when they saw her painted orange?\\
A9: they started laughing
\tcbline
Q10: Where did Cotton's mother put her to clean the paint off?\\
A10: a bucket of water
\tcbline
Q11: What did the other cats do when Cotton emerged from the bucket of water?\\
A11: licked her face
\tcbline
Q12: Did they want Cotton to change the color of her fur?\\
A12: no
\end{promptbox}

\begin{promptbox}[Test input with perturbation RS (with $EC(\cdot)$ result)]
\footnotesize\ttfamily
Q1: Whose paint was it? (True)
\tcbline
Q2: Was Cotton happy that she looked different than the rest of her family? (True)
\tcbline
Q3: Did she live alone? (True)
\tcbline
Q4: What did Cotton's mother and siblings do when they saw her painted orange? (True)
\tcbline
Q5: Where did Cotton's mother put her to clean the paint off? (True)
\tcbline
Q6: What did she do to try to make herself the same color as her sisters? (True)
\tcbline
Q7: Did they want Cotton to change the color of her fur? (True)
\tcbline
Q8: Who did she live with? (True)
\tcbline
Q9: What color were her sisters? (True)
\tcbline
Q10: What color was Cotton? (True)
\tcbline
Q11: Where did she live? (True)
\tcbline
Q12: What did the other cats do when Cotton emerged from the bucket of water? (True)
\end{promptbox}

\begin{promptbox}[Test input with perturbation RR (with $EC(\cdot)$ result)]
\footnotesize\ttfamily
Q1: What color was Cotton? (True)
\tcbline
Q2: Where did she live? (True)
\tcbline
Q3: Who did she live with? (True)
\tcbline
Q4: What color were her sisters? (True)
\tcbline
Q5: What did she do to try to make herself the same color as her sisters? (True)
\tcbline
Q6: Where did Cotton's mother put her to clean the paint off? (True)
\tcbline
Q7: What did the other cats do when Cotton emerged from the bucket of water? (True)
\tcbline
Q8: Did they want Cotton to change the color of her fur? (True)
\end{promptbox}

\begin{promptbox}[Test input with perturbation RD (with $EC(\cdot)$ result)]
\footnotesize\ttfamily
Q1: What color was Cotton? (True)
\tcbline
Q2: Where did she live? (True)
\tcbline
Q3: Did she live alone? (True)
\tcbline
Q4: What color was Cotton? (True)
\tcbline
Q5: Who did she live with? (True)
\tcbline
Q6: Where did she live? (True)
\tcbline
Q7: What color were her sisters? (True)
\tcbline
Q8: Was Cotton happy that she looked different than the rest of her family? (True)
\tcbline
Q9: What did she do to try to make herself the same color as her sisters? (True)
\tcbline
Q10: Whose paint was it? (True)
\tcbline
Q11: What did Cotton's mother and siblings do when they saw her painted orange? (True)
\tcbline
Q12: Where did Cotton's mother put her to clean the paint off? (True)
\tcbline
Q13: What did the other cats do when Cotton emerged from the bucket of water? (True)
\tcbline
Q14: What did the other cats do when Cotton emerged from the bucket of water? (True)
\tcbline
Q15: Did they want Cotton to change the color of her fur? (True)
\end{promptbox}

\begin{promptbox}[Test input with perturbation RSR (with $EC(\cdot)$ result)]
\footnotesize\ttfamily
Q1: What color were her sisters? (True)
\tcbline
Q2: What did she do to try to make herself the same color as her sisters? (True)
\tcbline
Q3: What did the other cats do when Cotton emerged from the bucket of water? (True)
\tcbline
Q4: Did they want Cotton to change the color of her fur? (True)
\tcbline
Q5: Who did she live with? (True)
\tcbline
Q6: Where did Cotton's mother put her to clean the paint off? (True)
\tcbline
Q7: What color was Cotton? (True)
\tcbline
Q8: Where did she live? (True)
\end{promptbox}

\begin{promptbox}[Test input with perturbation RSD (with $EC(\cdot)$ result)]
\footnotesize\ttfamily
Q1: What did she do to try to make herself the same color as her sisters? (True)
\tcbline
Q2: What did the other cats do when Cotton emerged from the bucket of water? (True)
\tcbline
Q3: Was Cotton happy that she looked different than the rest of her family? (False)
\tcbline
Q4: What color were her sisters? (True)
\tcbline
Q5: Did they want Cotton to change the color of her fur? (True)
\tcbline
Q6: What did the other cats do when Cotton emerged from the bucket of water? (True)
\tcbline
Q7: Where did she live? (True)
\tcbline
Q8: Did she live alone? (True)
\tcbline
Q9: Whose paint was it? (True)
\tcbline
Q10: What color was Cotton? (True)
\tcbline
Q11: Who did she live with? (True)
\tcbline
Q12: Where did Cotton's mother put her to clean the paint off? (True)
\tcbline
Q13: What color was Cotton? (True)
\tcbline
Q14: Where did she live? (True)
\tcbline
Q15: What did Cotton's mother and siblings do when they saw her painted orange? (True)
\end{promptbox}

The overall similarity for Q3 is 0.795 which is marginally below than the threshold (0.8), its context is mistakenly categorised as ``Inequivalent'' in this case. 

\begin{promptbox}[Test input with perturbation RRD (with $EC(\cdot)$ result)]
\footnotesize\ttfamily
Q1: What color was Cotton? (True)
\tcbline
Q2: Where did she live? (True)
\tcbline
Q3: Who did she live with? (True)
\tcbline
Q4: What color was Cotton? (True)
\tcbline
Q5: What color were her sisters? (True)
\tcbline
Q6: Where did she live? (True)
\tcbline
Q7: What did she do to try to make herself the same color as her sisters? (True)
\tcbline
Q8: Where did Cotton's mother put her to clean the paint off? (True)
\tcbline
Q9: What did the other cats do when Cotton emerged from the bucket of water? (True)
\tcbline
Q10: Did they want Cotton to change the color of her fur? (True)
\end{promptbox}

\begin{promptbox}[Test input with perturbation RSRD (with $EC(\cdot)$ result)]
\footnotesize\ttfamily
Q1: Where did Cotton's mother put her to clean the paint off? (True)
\tcbline
Q2: What color was Cotton? (True)
\tcbline
Q3: Who did she live with? (True)
\tcbline
Q4: What did the other cats do when Cotton emerged from the bucket of water? (True)
\tcbline
Q5: Where did she live? (True)
\tcbline
Q6: What did she do to try to make herself the same color as her sisters? (True)
\tcbline
Q7: Did they want Cotton to change the color of her fur? (True)
\tcbline
Q8: What color were her sisters? (True)
\tcbline
Q9: What color was Cotton? (True)
\tcbline
Q10: Where did she live? (True)
\end{promptbox}